\documentclass[12pt,letterpaper]{article}
\usepackage[margin=1in]{geometry}
\usepackage[max2]{authblk}
\usepackage{color,xcolor}
\usepackage[T1]{fontenc}
\usepackage{libertinus}

\usepackage{booktabs}
\usepackage{array}
\usepackage{graphicx}
\usepackage{amsmath}
\usepackage{amssymb}
\usepackage{stmaryrd}
\usepackage{fdsymbol}
\usepackage{subcaption}
\newcommand{\descr}[1]{\noindent\emph{#1}}

\usepackage[subpreambles=true]{standalone}
\usepackage{import}
\IfStandalone{}{
    \usepackage{siunitx}
    \sisetup{mode=text}
}

\begin{document}

\title{Platform architecture determines whether recommendation algorithms can shape information quality on social media}

\author[a,1]{Mohammad Hammas Saeed}
\author[a,1]{David A. Broniatowski}
\author[a,1]{Joseph Simons}
\author[a]{Erica Gralla}
\author[b]{Manan Suri}
\author[b,2]{Giovanni Luca Ciampaglia}
\affil[a]{George Washington University, Washington DC, USA}
\affil[b]{University of Maryland, College Park, MD, USA}
\affil[1]{\it\small M.H.S. (Author One), D.A.B. (Author Two), and J.S. (Author Three) contributed equally to this work.}
\affil[2]{\it\small E-mail: gciampag@umd.edu}
\maketitle

\begin{abstract}
Social media platforms shape public discourse through two fundamental design choices that naturally co-occur in any field investigation: platform architecture, which defines what types of actors exist and how they interact, and recommendation algorithm, which determines what content is surfaced to users. Using agent-based simulation, we orthogonally manipulate both factors, exploring four prototypical architectures -- tree (e.g., Reddit), layered hierarchy (e.g., Facebook), network (e.g., Twitter), and complete graph (e.g., TikTok) -- and two algorithms: chronological (LIFO) and popularity-based (Hot). Drawing on prior theory that identifies and ranks canonical system architectures in terms of their flexibility we hypothesize that algorithmic effects on information spread and quality should be largest on the most flexible platforms and smallest on the most constrained ones. We find strong confirmation of this prediction. On tree-like platforms like Reddit, the algorithm has no detectable effect on information spread and quality. On layered hierarchies and networks like Facebook and Twitter, respectively, the Hot algorithm has modest positive effects on both the spread of information and its quality. On complete structures like TikTok, the Hot algorithm leads to a winner-take-all dynamics that has strong negative effects on both information spread and quality, making the relation between content quality and popularity unpredictable. These findings imply that architectural considerations are more powerful levers than algorithmic interventions for the design of healthy online spaces and public discourse. Platform reform efforts focused exclusively on algorithm choice may be insufficient on architecturally unconstrained platforms and unnecessary on architecturally constrained ones.
\end{abstract}





\newpage

Social media platforms are used by billions of people worldwide for information access, networking, and civic participation. Their ability to disseminate content at scale with minimal centralized oversight has raised concerns about the spread of low-quality, misleading, and harmful information. These concerns have intensified as platforms have come to play a central role in public health communication, electoral discourse, and crisis response, calling for reform of these technologies~\cite{orben2025fixing}.

Two platform design choices are widely believed to shape these outcomes. The first is platform architecture -- the high-level design specification that defines, among others, (\emph{i}) what types of actors exist on a platform (e.g., individual users, moderators, pages, subreddits, and channels), (\emph{ii}) what types of connections are permitted among them (e.g., following, subscribing, group membership, and moderating), and (\emph{iii}) what actions each actor type is permitted to take over these connections. There is growing consensus among scholars and technologists that current platforms dot not fully explore the full design space of all possible specifications~\cite{jhaver2023decentralizing,zhang_form-_2024,li2024platform, Broniatowski_etal2023,broniatowski2025explaining,smaldino2025information}. 

The second design choice is the recommendation algorithm -- the ranking and filtering logic that determines which content, among all that is architecturally accessible to a user, is actually surfaced in their feed~\cite{eckles2022algorithmic}. There is an increasing body of evidence suggesting that algorithms shape key qualities of what content is amplified -- in particular its informativeness, emotional charge, and quality~\cite{ciampaglia_how_2018, huszar2022algorithmic, brady2023algorithm, wang_lower_2024, milli2025engagement, piccardi2025reranking}. Recent proposals for reform of social media are largely focusing on this latter aspect~\cite{moehring2025better}, by giving users and regulators greater control over recommendation algorithms -- for example, through algorithm auditing, increased algorithm choice, or the decentralized algorithm marketplaces being piloted by platforms such as Bluesky~\cite{kleppmann2024bluesky,ovadya2023bridging}. 

These proposals rest on an implicit assumption: that algorithms are the primary
lever through which platform design shapes information quality, and that
architectural differences among platforms are either less important or can be
held fixed while algorithms are varied. 
At the same time, theoretical frameworks have formalized how architectural differences among platforms can be measured and compared along a flexibility spectrum~\cite{broniatowski_measuring_2016, wei_characterizing_2019}. 
This prior work suggest that different platform architectures may constrain the efficacy of policy and algorithmic interventions by platforms in different ways. 

Retrospective analysis of the attempts made by Facebook to reduce the amount of
vaccine misinformation in circulation on its platform during the COVID-19
pandemic have shown that the architectural affordances of that platform (an
example of a `flexible layered' architecture; see Table~\ref{tab:diagram} for a
list of canonical architectures and their possible instantiations) allowed
motivated users to circumvent easily its content policies on vaccine
misinformation. As a result, vaccine-skeptical content became more
misinformative and more likely to appear in the newsfeeds of users despite
sustained removal efforts by the platform~\cite{Broniatowski_etal2023}.
Similarly, the architecture of Twitter (an example of a `network' architecture)
was associated with increased virality and decreased information quality among
vaccine-skeptical accounts following platform-wide content moderation
actions~\cite{broniatowski2025explaining}.

Despite growing regulatory and scholarly attention to these matters, the
relative importance of architecture versus algorithm in determining the
spread and quality of information on social media remains poorly understood.
This is in part because architecture and algorithm are generally confounded on
real platforms. The hierarchical architecture of, e.g., Reddit and its (until
recently default) chronological feed co-occur, as do the flat open-access
architecture of TikTok and its engagement-based
algorithm~\cite{Broniatowski_etal2023, broniatowski2025explaining}. This
co-occurrence makes it impossible to isolate the causal contribution of each
from observational data alone. It also limits the extent to which causal
evidence on the effects of both policy and algorithmic interventions from one
platform generalizes to different platform contexts. 

At present, no concrete system or testbed exists that would allow one to
manipulate in a systematic way both the architecture and algorithm of a real
social media platform. Nonetheless, if architecture is the primary driver of
either (or both) information spread and quality outcomes -- and if algorithmic
effects are contingent on architecture rather than independent of it -- then
reform efforts focused exclusively on algorithm choice may be insufficient, or
may produce qualitatively different effects on platforms with different
underlying architectures than those on which they were studied. Lacking
systematic ways to measure the impact of these two variables independently of
each other, we turn to computer simulations of a model social media system to at
least test the assumptions underpinning current attempts to reform social media
without the ambiguities of purely narrative explanations. 
 
To do so, we define and analyze counterfactual worlds populated by independent,
identically distributed populations of agents drawn from a common probability
distribution over agent attributes. We then orthogonally vary platform architectures
and recommendation algorithms across eight conditions defined by the factorial
combination of four architectures (Reddit, Facebook, Twitter, TikTok) and two
algorithms: a chronological algorithm in which agents review content in reverse
order of production (Last-In-First-Out or LIFO), which represents the
pre-personalization default on most platforms, and a popularity-based algorithm
in which content is ranked by accumulated engagement (Hot). Holding the
agent-generating process fixed across conditions ensures comparability and
isolates, within the context of our model, the main and interaction effects of
platform architecture and recommendation algorithm on our simulated outcomes. 

At every time step in our model, an agent is selected at random to become active. With some probability $p_{\rm post}$, an active agent can post a new message. Else, if the agent did not post a new message, with some other independent probability $p_{\rm share}$, the agent reviews all the messages in its feed (i.e., it is exposed to them) and decides whether to re-share one of them, as long as (\emph{i}) its quality exceeds the individual quality preference threshold of the agent and (\emph{ii}) its magnitude exceeds the agent's likelihood to share content. Finally, regardless of these previous actions, an active agent picks a message to like in its feed, chosen in proportion to its motivational appeal. Both of these forms of engagement (sharing and liking) are meant to capture individual differences in sensitivity to content quality. (See Supplementary Information for a full definition of agent and message characteristics, and for the simulation loop pseudocode.)

Content quality varies continuously from $-1$ (low quality) to $+1$ (high quality) and is a measure of its informativeness. Each message carries two additional properties that determine its strength or magnitude: whether it conveys a meaningful gist representation (illuminating power) and whether it is motivational (motivating valence). These choices are inspired by a prior model of information sharing whose predictions have been validated against observational data~\cite{broniatowski2019illuminate, broniatowski2016effective, broniatowski2024role}. These properties could allow us in principle to distinguish demand-side explanations for information, in which content quality and individual quality preferences determine what circulates, from supply-side explanations rooted in platform architecture and algorithm, which determine what content agents are exposed to in the first place. 

Agent population size, as well as the distributions of content quality and of agent quality preferences, are explicitly modeled as features of the simulation and held constant across conditions. This choice allows us to estimate, while controlling for individual differences in quality preference and content composition, the effect of algorithm and architecture on several outcomes related to the spread of messages. For each message we define its `reach' (i.e., the number of agents in whose feed the message was added, regardless of whether it was ultimately reviewed or not), its `exposure' (i.e., the number of agents who reviewed the message), and its `engagement' (i.e., the number of agents who reshared or liked the message). 

For each of these metrics, we further define across all algorithm--architecture combinations (\emph{i}) its ``breadth'' (i.e., the proportion of messages with non-zero value of the metric), (\emph{ii}) its ``depth'' (i.e., its overall value across all messages), and (\emph{iii}) its ``quality'' (i.e., the mean content quality of messages with non-zero values in that metric). Ten independent simulation runs with different random seeds provide estimates of cross-run variability and allow us to assess the robustness of findings to stochastic variation in agent behavior.

\section*{Results}
We ran each simulation for 10,000 time steps (i.e. agent activations) across eight conditions defined by the factorial combination of architecture (Reddit, Facebook, Twitter, or TikTok) and algorithm (LIFO or Hot), replicated across 10 simulation runs with different random seeds. Each simulation run yielded approximately 4,464 messages per condition (range: 4,359--4,561), for a total of 357,088 messages across all conditions. 

Moses' theory of generic architectures ranks architectures by flexibility -- the extent to which they impose constraints on how the system behaves~\cite{broniatowski_measuring_2016}. Applied to social media, it predicts that the effect of a recommendation algorithm on information spread and quality should be largest on the most architecturally flexible platforms and smallest on the most architecturally constrained ones (see Materials and Methods). 

We test this prediction by examining the contrast between the LIFO and Hot algorithm within each platform. We use hurdle models and combine both model components (see Materials and Methods) to estimate the mean exposures per message as the product of exposure breadth (the proportion of messages reviewed by at least one agent) and exposure depth (the mean number of agents who reviewed a message, among messages reviewed by at least one agent). A similar consideration can be done for message reach and engagement. These metrics capture different aspects of the total information flow across all messages and allow us to test the flexibility hypothesis of Moses' theory. 

\begin{figure}[t]
    \includegraphics[width=\linewidth]{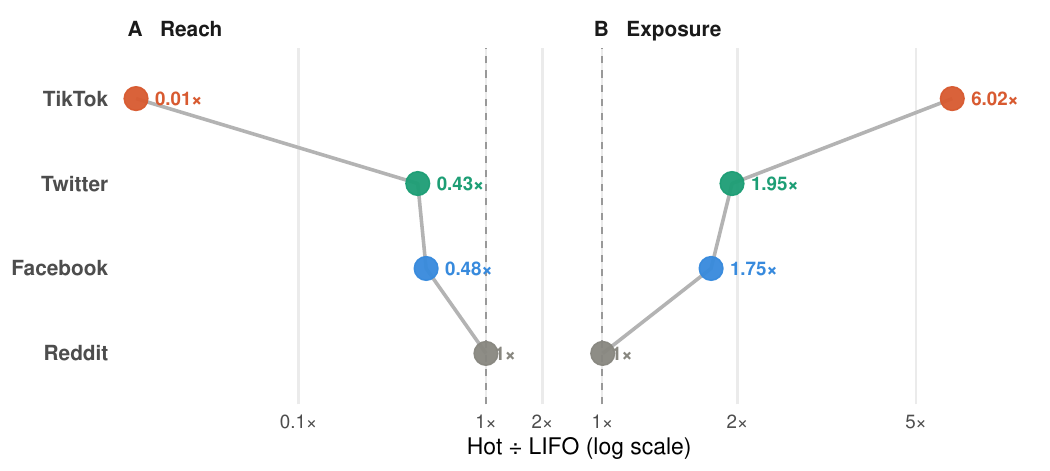}
    \caption{Hot $\div$ LIFO ratio of mean number of agents reached or exposed per message, in log scale. The dashed line corresponds to no effect (1$\times$ ratio). Platforms are ordered by Moses' flexibility metric: Reddit (tree) $<$ Facebook (layered) $<$ Twitter (network) $<$ TikTok (complete graph). \textbf{Panel A:} Reach; the Hot algorithm reduces the total number of agents reached per message monotonically with architectural flexibility, from no effect on Reddit to a 99\% reduction on TikTok. \textbf{Panel B:} Exposure; the Hot algorithm increases the total number of exposed agents each message in the same architectural order, from no effect on Reddit to a 6$\times$ increase on TikTok. The mirror-image ordering across panels confirms that architectural flexibility determines the magnitude of the algorithm's effect regardless of direction.}
\label{fig:algorithm_contrasts}
\end{figure}

Figure~\ref{fig:algorithm_contrasts} shows this ratio for both reach (Panel A) and exposure (Panel B). In both panels, the Hot/LIFO ratio is ordered from least to most flexible as predicted by Moses' theory, though in opposite directions: the Hot algorithm reduces message reach the most on TikTok (Panel A), while it increases message exposure the most on TikTok (Panel B). In either cases, Reddit is unchanged.

\begin{figure*}[t]
    \includegraphics[width=\textwidth]{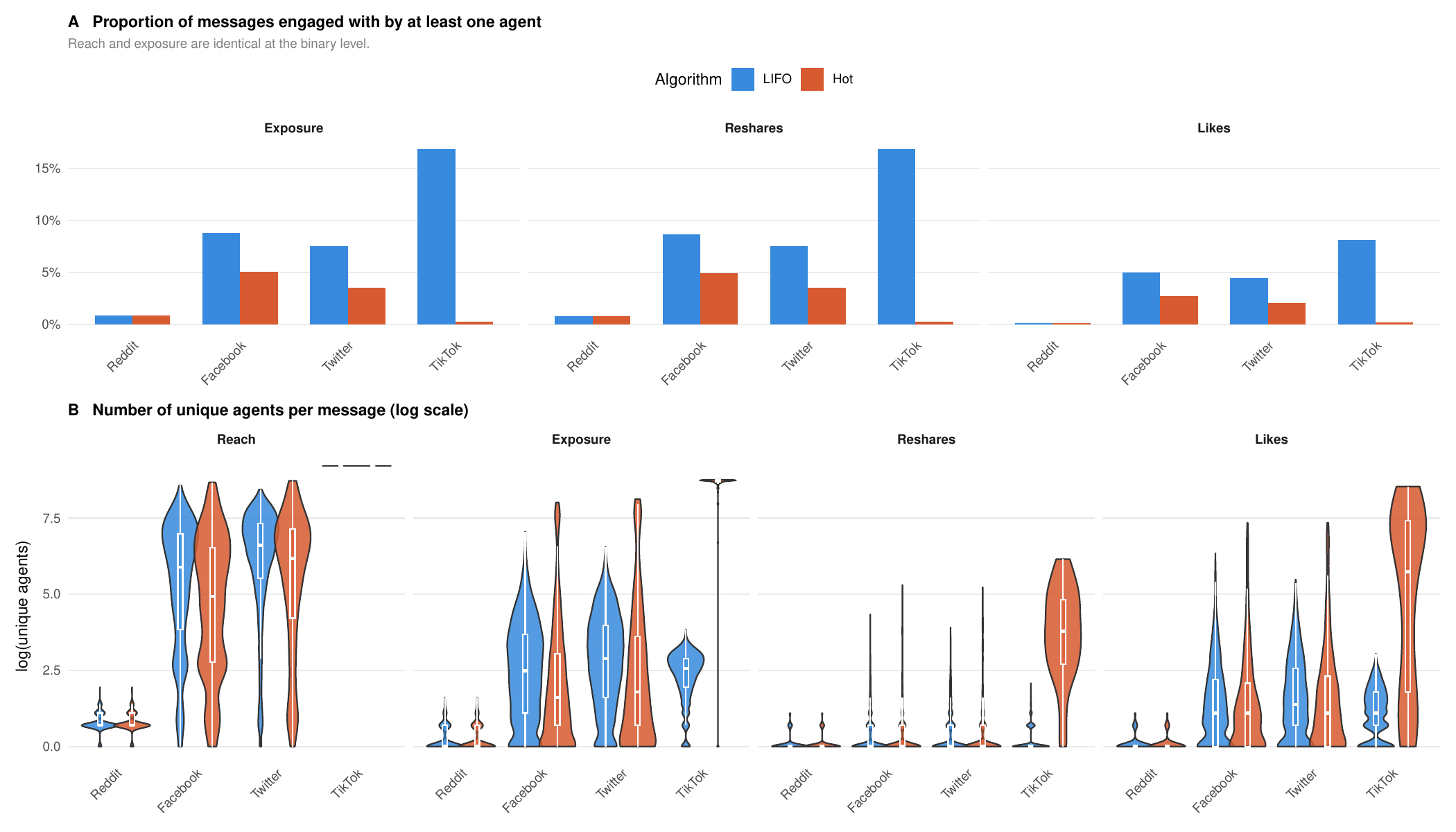}
    \caption{
    Panel A shows the breadth of exposure and engagement (broken down by engagement type); reach is not shown since it is identical to exposure. Panel B shows the depth of reach, exposure, and engagement (again broken down), in log scale.}
    \label{fig:violin_reshare}
\end{figure*}

Figure~\ref{fig:violin_reshare}(A) shows the breadth of exposure and engagement (broken down by reshares and likes) across our simulations, while Figure~\ref{fig:violin_reshare}(B) shows the depth distributions for reach, exposure, and engagements (again broken down by reshares and likes) at the end of our simulation run for both Hot and LIFO algorithm. 
Panel A omits reach, since reach and exposure are identical at the binary level; however the two diverge in count (Panel B). In other words, viral messages reach the feeds of many agents but not all agents review them before displacement. In fact, across all combinations, the majority of messages is never seen by any agent nor receives any engagement (Figure~\ref{fig:violin_reshare}(A)). 

In terms of reshares, we find that Reddit is the least affected by algorithm change where agent reshares follow similar distributions regardless of the choice of algorithm. Next, Facebook shows a greater variation highlighting the influence of friends, group and page subscriptions on platform flexibility. Similar to Facebook, Twitter shows a slightly higher change in agent reshares from changes in the algorithm since its architecture is more flexible. Lastly, TikTok shows the greatest variability and overall impact of the algorithm. Likes follow a qualitatively similar pattern to reshares.

\subsection*{Reddit: The Algorithm Has No Observable Effect}

On Reddit, changing from the LIFO algorithm to the Hot algorithm had no detectable effect on any outcome. Exposure breadth was identical under both algorithms (0.8\% of messages received any exposure by at least one agent), as was resharing breadth (0.8\%) and liking breadth (0.1\%). Depth was similarly unaffected: under both LIFO and Hot, the per message geometric means for agents reached, agents exposed, and agents who reshared, or agents who liked were 2.19, 1.28, 1.09, and 1.10, respectively. All six hurdle model algorithm effects on Reddit were indistinguishable from zero (all $p$ = 1.000), and all algorithm × Reddit interaction terms were similarly null. Exposure-weighted mean content quality was 0.364 for both algorithms. 

\subsection*{Facebook: Modest Algorithm Effects}
On Facebook, switching from LIFO to Hot reduced exposure breadth from 8.8\% to 5.0\% (OR = 0.54, $p < .001$), resharing breadth from 8.7\% to 4.9\% (OR = 0.54, $p < .001$), and liking breadth from 5.0\% to 2.7\% (OR = 0.51, $p$ = .001). For depth, Hot reduced mean agents reached per message from 683.4 to 573.6, and increased mean agents exposed, agents resharing, and agents liked, from 37.7 to 115.2,  2.08 to 4.02, and 11.15 to 29.72, respectively. In contrast, per message geometric means for agents reached and agents exposed declined from 11.6 to 7.46 (OR = 0.53, $p < .001$) and 11.6 to 7.46 (OR = 0.64, $p < .001$), respectively, reflecting increased concentration of exposure among a smaller number of messages. The Hot algorithm modestly improved information quality (exposure-weighted mean content quality increased from 0.409 to 0.481). 

\subsection*{Twitter: Modest Algorithm Effects}
On Twitter, switching from LIFO to hot reduced exposure breadth from 7.5\% to 3.5\% (OR = 0.44, $p < .001$), resharing breadth from 7.5\% to 3.5\% (OR = 0.44, $p < .001$), and liking breadth from 4.4\% to 2.0\% (OR = 0.44, $p < .001$).  For depth, Hot reduced mean agents reached per message from 986.5 to 920.5, and increased mean agents exposed, agents resharing, and agents liked, from 43.0 to 180.2,  2.24 to 5.28, and  12.27 to 40.97, respectively. In contrast, per message geometric means for agents reached and agents exposed declined from 432.7 to 229.6 (OR = 0.53, $p < .001$) and 16.0 to 10.1 (OR = 0.63, $p < .001$), respectively, again reflecting increased concentration of exposure among a smaller number of messages. The Hot algorithm modestly improved information quality (exposure-weighted mean content quality increased from 0.423 to 0.482). 

Twitter produced greater depth than Facebook across all conditions (see Supplementary Materials). 

\subsection*{TikTok: The Algorithm Is Focal}
On TikTok, the Hot algorithm produced dramatic effects on all outcomes. Exposure breadth decreased from 16.9\% under LIFO -- the highest of any condition -- to 0.2\% under Hot -- the lowest of any condition (OR = 0.01, $p < .001$). Resharing breadth also decreased (from 16.9\% to 0.2\%, OR = 0.01, $p < .001$) and liking breadth from 8.1\% to 0.1\% (OR = 0.02, $p < .001$). 

The fully-connected structure of TikTok ensured that messages, once reaching a feed, were reviewed by virtually all agents regardless of algorithm (OR = 0.05, $p = 0.78$) However, once received, all other measures of depth increased under Hot compared to LIFO: the geometric mean number of agents exposed, resharing, and liking each message increased from 10.2 to 4,563 (OR = 477.3, $p < .001$), 1.21 to 39.0 (OR = 35.7, $p < .001$), and 3.02 to 128.9 (OR = 54.1, $p < .001$), respectively. 

Information quality decreased correspondingly. Exposure-weighted mean content quality fell from 0.382 under LIFO to 0.032 under Hot -- a 92\% reduction. Mixed-effect models of content quality (see Materials and Methods) suggest that 91\% of this collapse was a direct effect of the algorithm operating within TikTok's unconstrained architecture, not an artifact of differential surfacing of content types ($b$ = $-0.305$ vs $-0.336$ after controlling for content composition, $p < .001$). 

The quality of the locked-in content was not systematically low; rather, it was random. Across ten independent simulation runs with identical parameters, exposure-weighted content quality ranged from $-0.243$  to $+0.439$, a range of 0.682, compared to a range of only 0.083 across seeds for TikTok/LIFO (see Figure~\ref{fig:lollipop}). 

\begin{figure}[t] 
    \includegraphics[width=\linewidth]{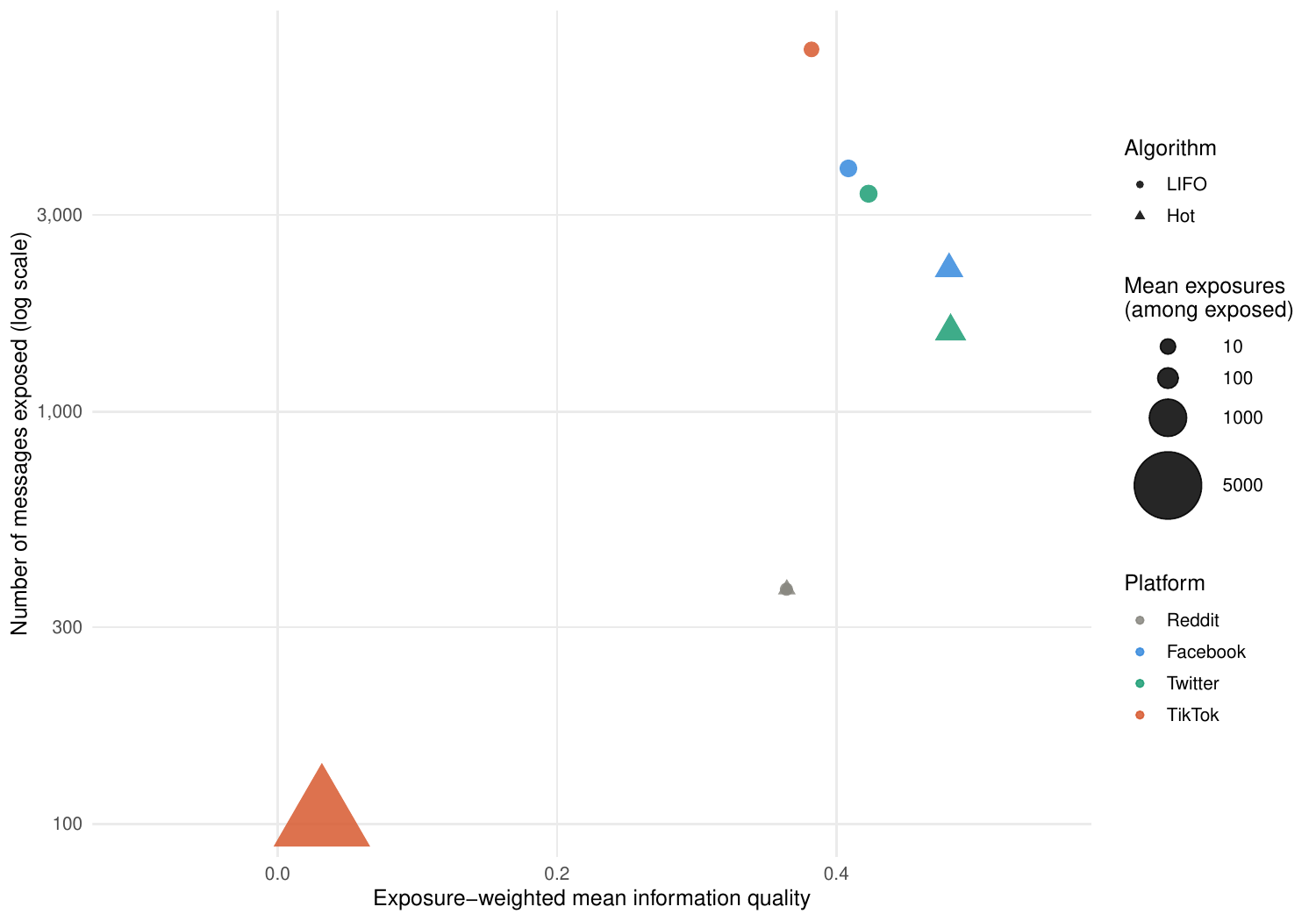}
    \caption{Quality, breadth, and depth of information exposure by platform and algorithm. Each bubble represents one platform $\times$ algorithm condition. X axis: exposure-weighted mean content quality (linear scale, depth). Y axis: number of messages receiving any exposure (log scale, breadth). Marker size is proportional to mean exposures per message among exposed messages (depth). Platform indicated by color; algorithm by marker shape (circle = LIFO, triangle = Hot).}
    \label{fig:bubble}
\end{figure}

\subsection*{Information Quality, Breadth, and Depth Across Conditions}

Figure~\ref{fig:bubble} summarizes how information quality, breadth, and depth vary across all eight conditions. Under LIFO, all platforms produced positive exposure-weighted content quality (range: 0.364--0.423), with Reddit showing the lowest breadth and depth and TikTok the highest breadth. The Hot algorithm had negligible effects on Reddit, modest positive effects on Facebook and Twitter quality, and larger negative effects on TikTok quality. In particular, the TikTok/Hot combination moves sharply left (quality collapse), downward (breadth collapse), and increases dramatically in size (depth explosion), reflecting winner-take-all lock-in.

\begin{figure*}[t] 
    \includegraphics[width=\textwidth]{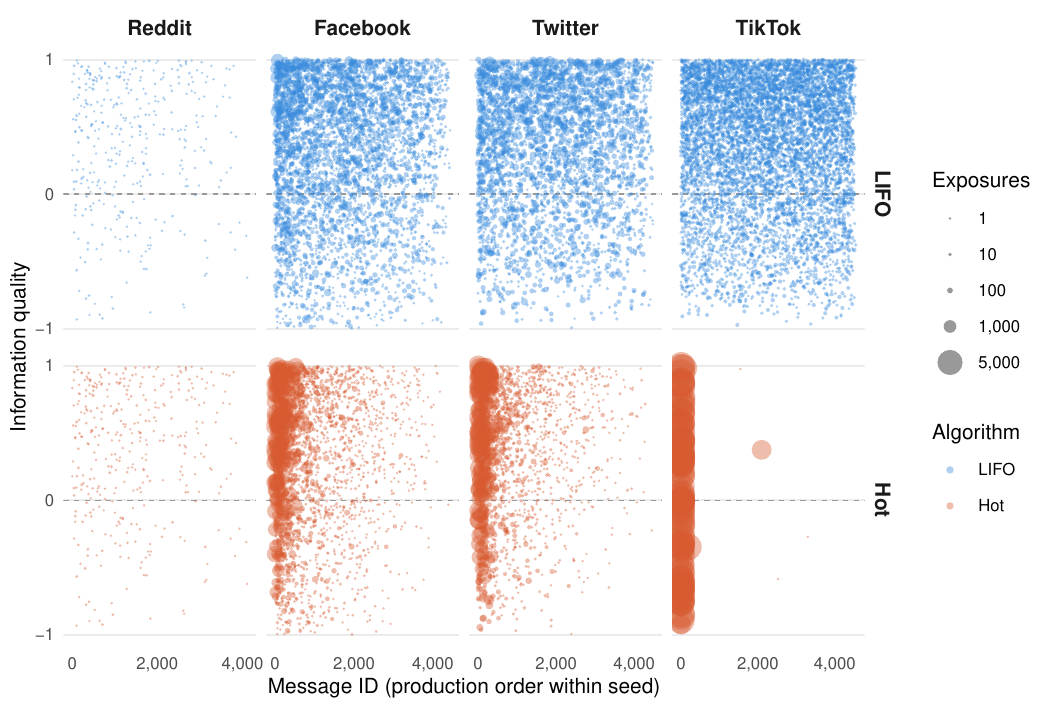}
    \caption{Content quality and production order of exposed messages for all architectures under LIFO and Hot algorithms, combining data from multiple simulation runs together. Each circle represents one message receiving any exposure. The X axis is the message ID (i.e., its production order within each run). The Y axis the message quality. Circle size is proportional to the number of exposures received. Color: algorithm (blue = LIFO, orange = Hot).}
    \label{fig:exposure_scatter}
\end{figure*}

This interpretation is confirmed by Figure~\ref{fig:exposure_scatter}, that reveals that under TikTok/Hot, exposed messages were almost entirely restricted to those produced in the earliest simulation time steps, with larger exposure counts, confirming that lock-in selected on timing rather than quality. In contrast, Reddit shows sparse, uniformly small exposure counts regardless of algorithm. 

\begin{figure*}
    \includegraphics[width=0.68\textwidth]{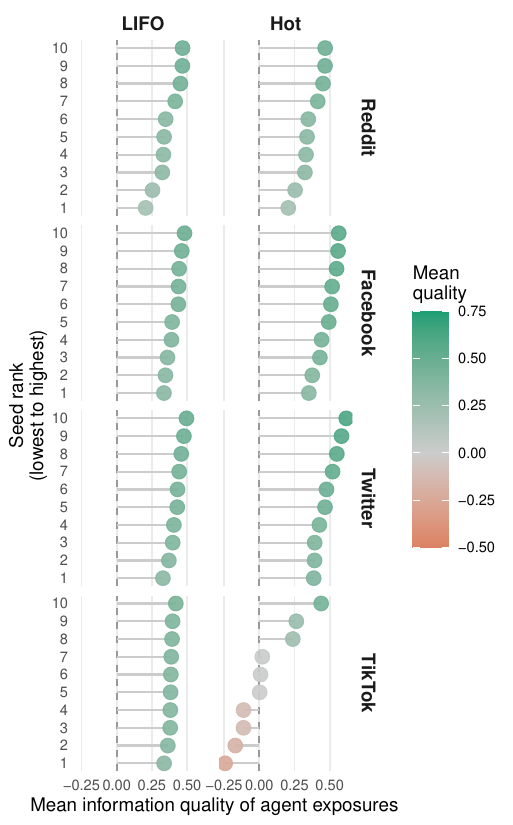}
    \caption{Seed-level mean information quality of agent exposures by platform $\times$ algorithm condition. Each point is one simulation seed, ranked from lowest (1) to highest (10) exposure-weighted mean quality within each condition. Horizontal segments connect each seed's mean quality to zero. The color of the marker is mapped to quality.}
    \label{fig:lollipop}
\end{figure*}

Finally, analysis of the individual random seeds used across simulation runs in Figure~\ref{fig:lollipop} shows that quality varied substantially across seeds only in the TikTok/Hot combination, whereas it was consistent across seeds under all other conditions. This confirms that the winner-take-all mechanism is specific to the combination of an unconstrained architecture and a popularity-based algorithm.

\section*{Discussion}

Our findings demonstrate that platform architecture strongly determines how information flows on social media platforms. In contrast, recommendation algorithm effects appear to be conditioned on the architectural context in which they operate. This pattern is the empirical signature of the flexibility of generic architectures predicted by Moses' theory~\cite{broniatowski_measuring_2016}: more flexible architectures impose fewer constraints on how the algorithm shapes information flow. The mirror-image Moses ordering of the two ratios in Fig.~\ref{fig:algorithm_contrasts} -- reach declining most on the most flexible platform but exposure increasing most on the most flexible platform -- confirms that architectural flexibility determines the magnitude of the algorithm's effect on both reach and exposure simultaneously, but in opposite directions. The result is a massive concentration of exposure on a small subset of messages that are able to reach the feeds of any agent.

On Reddit, the Hot algorithm had essentially no effect across all outcomes, confirming that the presence of subreddits, acting as collectors of messages, behave as a sort of `structural gate' that suppresses the impact of the algorithm on individual user feeds. Although our model did not explicitly include moderator agents in these subreddits, in the real world the actions of human moderators are likely to strengthen this filtering effect. 

In contrast, on TikTok, the Hot algorithm had catastrophic effects on breadth, depth, and quality, confirming that TikTok's complete graph, which imposes no structural constraints on information flow, gives the algorithm maximum freedom to determine what spreads and to whom.

The differences between reach and exposure on TikTok under Hot illustrates the winner-take-all mechanism at play. The Hot algorithm did not cause viral messages to reach more agents. Rather, agents' exposures were almost exclusively focused on the small number of messages that entered circulation. Winner-take-all lock-in on TikTok is therefore not a property of how widely content spreads through the network, but of how the Hot algorithm privileges a small number of messages for exposure. These messages were the ones produced in the earliest time steps of the simulation. The result is that early winners monopolize agent queues, accumulating enormous exposure counts while crowding out all subsequently produced content. 

As a result, the quality of this locked-in content is a function of the initial conditions of the system, and specifically, of the messages that happen to be produced before agent queues fill to capacity. Our results suggest that the dynamics of highly flexible systems like TikTok may exhibit behaviors resembling a tipping point. Even when a majority of the content is high quality, a relatively small amount of viral, low quality information can dominate users' queues relatively quickly if it is introduced early on. The TikTok/Hot condition therefore represents a qualitatively different regime from all other conditions. 

Overall, these findings suggest a possible reframing of the TikTok misinformation problem: the issue is not that TikTok's algorithm systematically selects low-quality content, but that it amplifies whatever content happens to achieve early traction regardless of quality -- a structural lottery whose outcomes are unpredictable by design, yet are a function of both content quality and user demand, which may prime the system for an information cascade of low (or high) quality content. This is consistent with the empirical finding that content moderation interventions on architecturally flexible platforms tend to be ineffective or counterproductive~\cite{Broniatowski_etal2023, broniatowski2025explaining}: if the information environment is determined by early lock-in rather than by sustained quality-based selection, then post-hoc removal of low-quality content will simply be replaced by whichever content happens to fill the resulting vacancy. These results may also explain recent counterintuitive empirical findings ~\cite{Broniatowski2022COVIDMisinformation} that social media posts about COVID-19 contained proportionally less misinformation than misinformation about other topics and in prior years, despite a widespread perception that misinformation was ubiquitous. 

Our observation that the Hot algorithm modestly improves information quality on Facebook and Twitter is also counterintuitive in light of the broader literature associating engagement-based ranking with misinformation amplification~\cite{ciampaglia_how_2018, van_der_linden_how_2024}. We interpret this as a consequence of intermediate architectural flexibility: on structured platforms, content must pass through quality-sensitive resharing chains before accumulating engagement. Agents on these platforms only reshare content that exceeds their quality preference threshold, so the Hot algorithm amplifies content that has already been filtered for quality. This interpretation predicts that the positive quality effect of Hot should diminish or reverse as platform architecture becomes less constrained, which is exactly the pattern observed on TikTok. This finding is consistent with Wang et al.~\cite{wang_lower_2024}, who find that algorithmic timelines on Twitter surface more reliable content than chronological feeds, and with Baumann et al.~\cite{baumann_optimal_2024}, who identify conditions under which engagement optimization and content quality are compatible rather than opposed.

In addition, both Moses' theory and prior empirical
findings~\cite{Broniatowski_etal2023,broniatowski2025explaining} predict that
Twitter's network architecture should be more flexible than Facebook's layered
hierarchy, producing larger algorithm
effects~\cite{broniatowski_measuring_2016}. Our results are directionally
consistent with this prediction: Twitter showed greater depth than Facebook
under both algorithms across all three outcomes, and the breadth reduction under
Hot was slightly larger on Twitter (56\% relative decline) than Facebook (43\%).

However, the difference between the two platforms was modest, and no clear
ordering emerged for the algorithm contrast itself. The failure to observe a
clearer separation between Facebook and Twitter is likely a consequence of
simulation scale rather than a theoretical failure. Moses' flexibility metric
predicts that the difference between a layered hierarchy and a network
architecture grows with the number of nodes -- their orders of growth are within
the same asymptotic class for small systems~\cite{broniatowski_measuring_2016}.
With a fixed agent population that uses similar implementations of Facebook and
Twitter user networks (see Materials and Methods) our simulations may not have been large enough to produce
the separation that the theory predicts at scale. Future work with larger
simulations should be able to resolve this ordering empirically.

These findings have direct implications for the reform proposals discussed in the Introduction~\cite{moehring2025better}. Given that our findings show whether algorithm choice has any effect at all is dependent on the choice of platform architecture, platform governance efforts that focus exclusively on algorithm choice, such as Bluesky's algorithm marketplace~\cite{kleppmann2024bluesky} or proposals for algorithm auditing~\cite{ovadya2023bridging} may be insufficient on architecturally unconstrained platforms and unnecessary on architecturally constrained ones. On Reddit-like architectures, algorithm choice may not matter; on TikTok-like architectures, algorithm choice fully determines whether the lock-in dynamic dominates. The actionable implication is that architectural constraints may be more powerful levers for information quality governance than algorithmic interventions, and should be the primary target of attempts to govern social media platforms.

Several limitations of this study warrant acknowledgment. First, our simulations model a simplified version of each platform architecture, and we hold agent population, content quality distributions, and quality preferences constant across conditions. Second, we compare only two algorithms, LIFO and Hot, covering only part of the algorithm design space. Real platforms use more complex ranking functions that combine recency, engagement, personalization, and other signals; the effects of these hybrid algorithms on architecturally constrained versus unconstrained platforms remain to be studied. Third, the agent quality preference parameter is drawn from a uniform distribution and held fixed over the course of the simulation; real users may update their preferences dynamically in response to the information environment they encounter, a feedback process not captured here.

Fourth, the operationalization of depth for the Moses ordering comparison can yield different ordinal rankings across conditions depending on whether arithmetic or geometric means are used. The arithmetic mean is sensitive to extreme values -- a small number of messages receiving very large engagement counts pulls the mean upward substantially -- while the geometric mean, which corresponds directly to the log-normal model coefficients, reflects the typical message rather than the average message and is therefore more robust to outliers. In the Supplementary Information, we report full orderings for both means, and although the ordering holds for the arithmetic mean, it does not always hold for the geometric mean. We do not view this as a failure of the Moses ordering prediction, but rather as a feature of it: the greater concentration of engagement on a small number of messages is itself a consequence of architectural flexibility, which removes structural constraints that would otherwise distribute engagement more evenly across messages. The arithmetic mean captures this concentration effect directly, while the geometric mean averages it away. Future work should develop more formal tests of the Moses ordering that account for the full shape of the depth distribution rather than relying on a single summary statistic. 

Finally, our model does not incorporate platform-wide content moderation policies or other similar platform-level interventions~\cite{coleman_combining_2022}. Findings on the efficacy of these interventions is mixed. Differences in architecture may explain these divergent findings.~\cite{Broniatowski_etal2023,broniatowski2025explaining,cima2025investigating, mccabe2024deplatforming}. Similarly, we do not explicitly model the role of volunteer moderators or other middle-level layers of governance distinct from platform ones~\cite{jhaver2023decentralizing}. Such examinations, and analysis of other platforms governance arrangements~\cite{cao-etal-2024-toxicity,jhaver2023decentralizing} are left to future work.

In conclusion, our results show that the effect of the choice of algorithm on information flow strongly depends on the choice of architecture. On highly-constrained platforms such as Reddit, the recommendation algorithm has no effect at all. On a fully connected platform like TikTok, the algorithm is focal because the architecture imposes no constraints on where information can go. This suggests that architecture imposes a fundamental constraint on what algorithmic governance can achieve, and that efforts to improve the quality of online information environments that focus exclusively on algorithms while leaving architecture unexamined may target the lever with the least purchase while leaving the largest determinant of quality untouched.


\section*{Materials and Methods}
\label{sec:methods}


\subsection*{Simulation Design}
\label{sec:simulation-design}

\begin{table*}
\centering
\caption{Summary of the architectures in our model.}
\small
\begin{tabular}{rm{.75in}p{1.25in}p{1.25in}p{1.5in}}
    \toprule
    {\bf Name} & {\bf Topology} & {\bf Nodes} & {\bf Edges} & {\bf Example platforms} \\
    \midrule
    Network & 
    \begin{minipage}{.75in}
      \includegraphics[width=.75in]{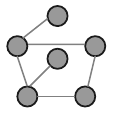}
    \end{minipage}
    & Users (circles $\textcolor{gray}{\medblackcircle}$) & user $\leftrightarrow$ user & Twitter/X, Weibo, Tumblr, Threads, Substack Notes, Bluesky \\
    
    Team / Complete & 
    \begin{minipage}{.75in}
      \includegraphics[width=.75in]{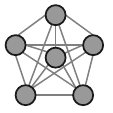}
    \end{minipage}
    & Users (circles $\textcolor{gray}{\medblackcircle}$) & user $\leftrightarrow$ user & TikTok, Instagram Reels, YouTube Shorts \\ 
    
    Tree & 
    \begin{minipage}{.75in}
      \includegraphics[width=.75in]{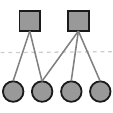}
    \end{minipage}
    & Users (circles $\textcolor{gray}{\medblackcircle}$), groups (squares $\textcolor{gray}{\medblacksquare}$) & user $\leftrightarrow$ community & Reddit, Usenet, Nextdoor, Web Forums \\ 
    
    Layered & 
    \begin{minipage}{.75in}
      \includegraphics[width=.75in]{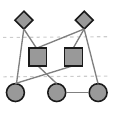}
    \end{minipage}
    & Users (circles $\textcolor{gray}{\medblackcircle}$), groups (squares $\textcolor{gray}{\medblacksquare}$), pages (diamonds $\textcolor{gray}{\medblackdiamond}$) & user $\leftrightarrow$ group, user $\leftrightarrow$ page, user $\leftrightarrow$ user, page $\leftrightarrow$ group, page $\leftrightarrow$ page & Facebook, VKontakte \\  
    \bottomrule
\end{tabular}
\label{tab:diagram}
\end{table*}

We construct a simulation framework that captures key features of multiple platform architectures and algorithms, and that features a parsimonious set of agent-level characteristics. In particular, we consider four prototypical architectures as stand-ins for four major classes of social media platforms (see Table~\ref{tab:diagram}): Network (inspired by flat microblogging platforms like Twitter/X), Team or Complete (for fully algorithmic platforms like TikTok), Tree (for community aggregator platforms like Reddit), and Layered (for social networks with different classes of nodes possessing different message transmission affordances, such as Facebook's users, groups, and pages). This choice of architectures deliberately captures the full range of the flexibility  spectrum~\cite{Broniatowski_Moses_2014}, while allowing us to study their interaction with different types of algorithms in common use by social media companies. Of course our model is necessarily a simplified description of the real-world counterparts of these architectures, which are constantly evolving and competing with each other by adopting common features and designs. Our goal is not to attain a realistic description of these systems, but rather to allow for controlled comparisons across scenarios while capturing key features of these real-world systems. 


\subsubsection*{Platform Architectures}

We first describe the four architectures from the most to the least flexible according to Moses' theory.

\descr{Team or Complete (e.g. TikTok).} TikTok, as well as specific features of other platforms like Instagram Reels or  YouTube Shorts, operates as a ``commons''\cite{zhang_form-_2024} -- i.e., a fully-connected user ``team'' or ``complete'' architecture, where the inventory from which recommendations are drawn encompasses the content created by any users on the platform (i.e., out-of-network recommendations), and not just the immediate neighbors of a user (in-network recommendations). This means that content propagation is not limited by prior relationships but is determined dynamically by engagement patterns. According to Moses' theory, a ``team'' should be the most flexible
and least controllable of the architectures included in this study.\cite{broniatowski_measuring_2016}

\descr{Network (e.g. Twitter/X).} A ``network'' is a flat social network architecture organized through a  follower model, as first popularized by microblogging platform Twitter (now X), Weibo, and Tumblr, and more recently by Threads, Substack Notes, and Bluesky. Users follow other users to see their content in an ordered timeline. Unlike the ``tree'' architecture (exemplified by Reddit, see below), there are no hard community boundaries, and unlike the ``layered'' architecture (exemplified by Facebook, see below), there is only one entity type. All users exist in a public, shared space, and content can traverse widely via resharing or retweeting. According to Moses' theory, a ``network'' is expected to be flexible but not very controllable compared to most of the platforms in this study, with the notable exception of the ``team''.\cite{broniatowski_measuring_2016}

\descr{Layered (e.g. Facebook).}
Mature social networking services like Facebook or VKontakte typically exhibit a multi-layered hierarchical architecture.\cite{Broniatowski_etal2023} The primary layers include users, where individual accounts can produce, share, and engage with content. Next is the layer for groups which are interest-based collectives where membership gates access and visibility, but all members can post and share. Lastly, and unlike the ``tree'', there is an additional
layer for pages (e.g., for brands or influencers) that broadcast to large audiences. Links in this system are therefore heterogeneous, denoting relationships such as friendship, group memberships, and page subscriptions. According to Moses' theory, the ``layered'' architecture should be moderately flexible and moderately controllable compared to the other architectures considered in this study.\cite{broniatowski_measuring_2016}

\descr{Tree (e.g. Reddit).} Platforms organized into topically bounded communities (groups, channels, etc.) can be thought as roughly resembling a tree-like structure (e.g. Nextdoor, where a house can only be part of one neighborhood) or, more commonly, a bipartite structure (e.g. Reddit, Usenet, or most web forums). According to Moses' theory, a ``tree'' is  the least flexible of the architectures considered in this study. Although platforms such as Reddit allow users to participate across groups (therefore violating a strict tree structure) content is still localized within groups by design. 


\subsubsection*{Recommendation Algorithms}
Next, we consider two major ranking algorithms in our simulations, corresponding to widely used ranking methods: reverse chronological, and popularity-based ranking. 

\descr{LIFO (Last-In, First-Out).} The LIFO algorithm prioritizes the most recently received message. This mechanism emphasizes recency and is often used in real-time communication platforms (e.g., messaging apps, live updates, and reverse chronological feeds). In the context of content feeds, LIFO is expected to lead to short-term attention dominance, where newer messages quickly displace older ones regardless of quality or relevance.

\descr{Hot.} The ``Hot'' algorithm ranks and selects the message with the highest number of reshares, approximating a dynamic popularity score, making popular messages dominate the feed. In the context of content feeds, Hot is expected to lead to self-reinforcement or rich-get-richer patterns, where messages that accrue early popularity displace news ones regardless of quality or relevance.


\subsection*{Model Mechanics}
Our model describes a generic social media platform as a system where messages are passed around different types of entities (users, groups, and pages). We refer to them as \emph{agents}, each belonging to a particular \emph{layer}, which defines the patterns of connectivity with the rest of the system (see Table~\ref{tab:diagram}). Common to all architectures is a layer of \emph{users}, who are capable of producing new messages and resharing messages they receive. Each user is equipped with two finite-size queues: an \emph{input queue} represents their news feed, and an \emph{output queue} represents their user profile, which collects all the messages posted or reshared by the user. When a user posts a message into their output queue, it is automatically pushed into the input queues of (\emph{i}) all others users (``team''), (\emph{ii}) users who either follow (``network'') or are friends (``layered'') with the user, or (\emph{iii}) in the queues of the groups the user is member of (``tree'' and ``layered''). In our model users behave following a model of human decision making inspired by Fuzzy-Trace Theory when deciding which messages to reshare or  like.~\cite{broniatowski2019illuminate} For full detail on characteristics of users and messages, see Supplementary Information.  

In addition to users, our model features two other types of agents: groups and pages (see Table~\ref{tab:diagram}). Their presence depends on the architecture: the ``network'' and the ``team'' only feature users; the `tree' and ``layered'' architectures both feature groups; and the ``layered'' architecture is the only one that features pages. 

In both the ``tree'' and ``layered'' architectures users are members of multiple groups. These are modeled as pure message repositories, each of which are associated with a single queue that acts simultaneously as both the input and output queue. When a user activates to perform an action (see below), they \emph{pull} all messages from the queues of the groups they are member of into their own input queue (which, in the ``layered'' architecture may already contain messages pushed from other users). If a user decides to post into a group, their message is added to that same group's queue.

Pages, on the other hand, behave as a hybrid between a group and a user. Like a user, they have both an input and output queue. As on Facebook, pages can follow other pages. When a page posts a message (either by creating a new message or by resharing a message from another page) into their output queue, it is automatically copied into the input queue of all pages that follow it. Like groups, users pull from the output queues of pages they follow when they activate to read their own feed. Finally, as on Facebook, pages can also be ``admins'' of groups. Thus, when a message is posted into the output queue of a page, it is automatically copied into the queues of all the groups it is admin of. 

For details about the sampling procedure to obtain simulations, see Supplementary Information.


\subsection*{Simulation}
We ran our simulations with 10,000 users over 10,000 time steps replicated over 10 seed values. To ensure that the parameters we use are representative, we calibrated them against empirical findings. Specifically, the \emph{post probability} is set to 0.45, reflecting observed activity rates in Twitter diffusion networks~\cite{weng2012competition}. The \emph{reshare probability} is set to 0.25, consistent with previously reported reshare behaviors~\cite{sasahara_social_2021}. The \emph{queue size}, representing an agent's screen or attention span, was set to 10, aligning with the previously reported default values~\cite{sasahara_social_2021}.

\begin{figure*}[t] 
    \centering
    \includegraphics[width=0.23\textwidth]{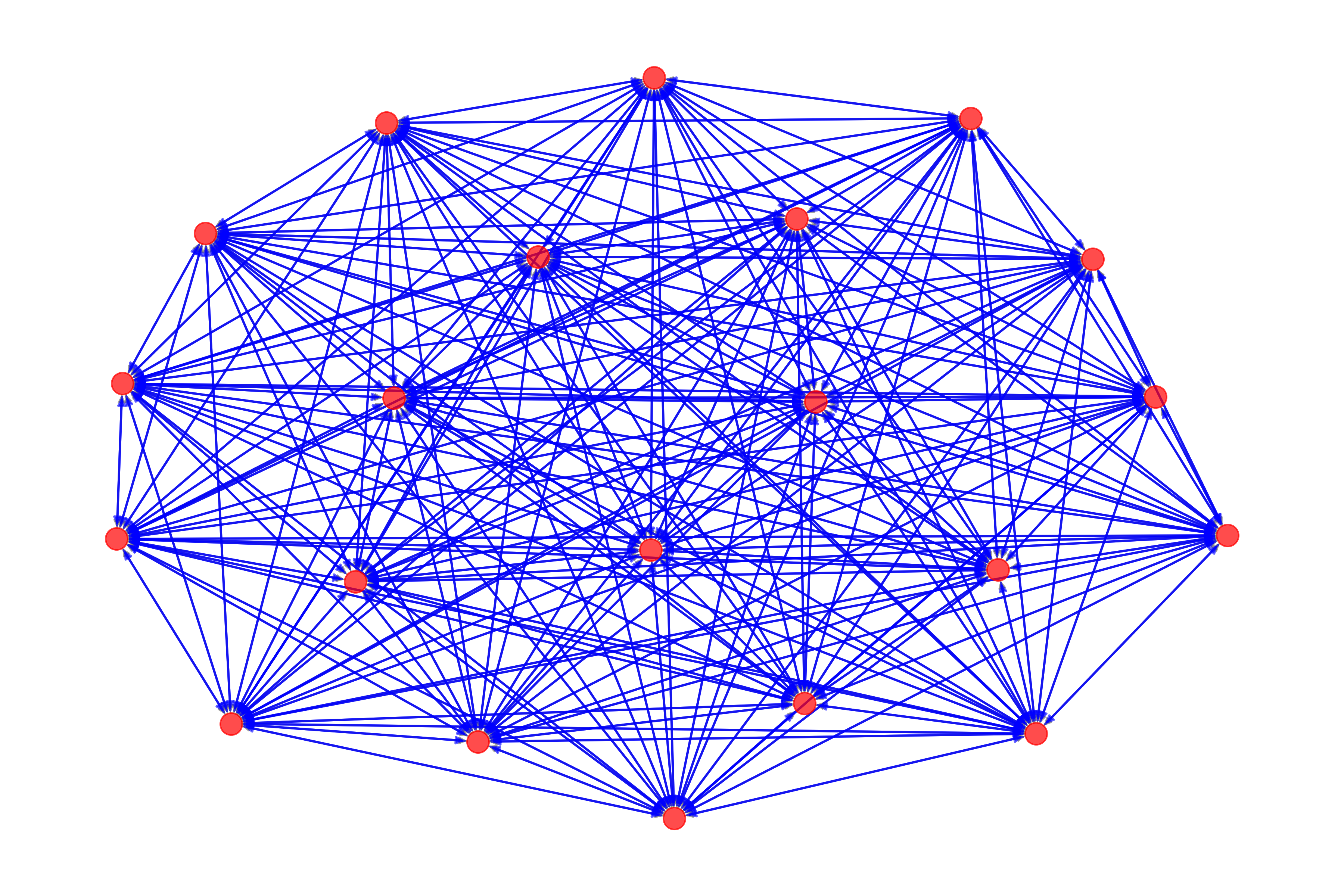}
    \hfill
    \includegraphics[width=0.23\textwidth]{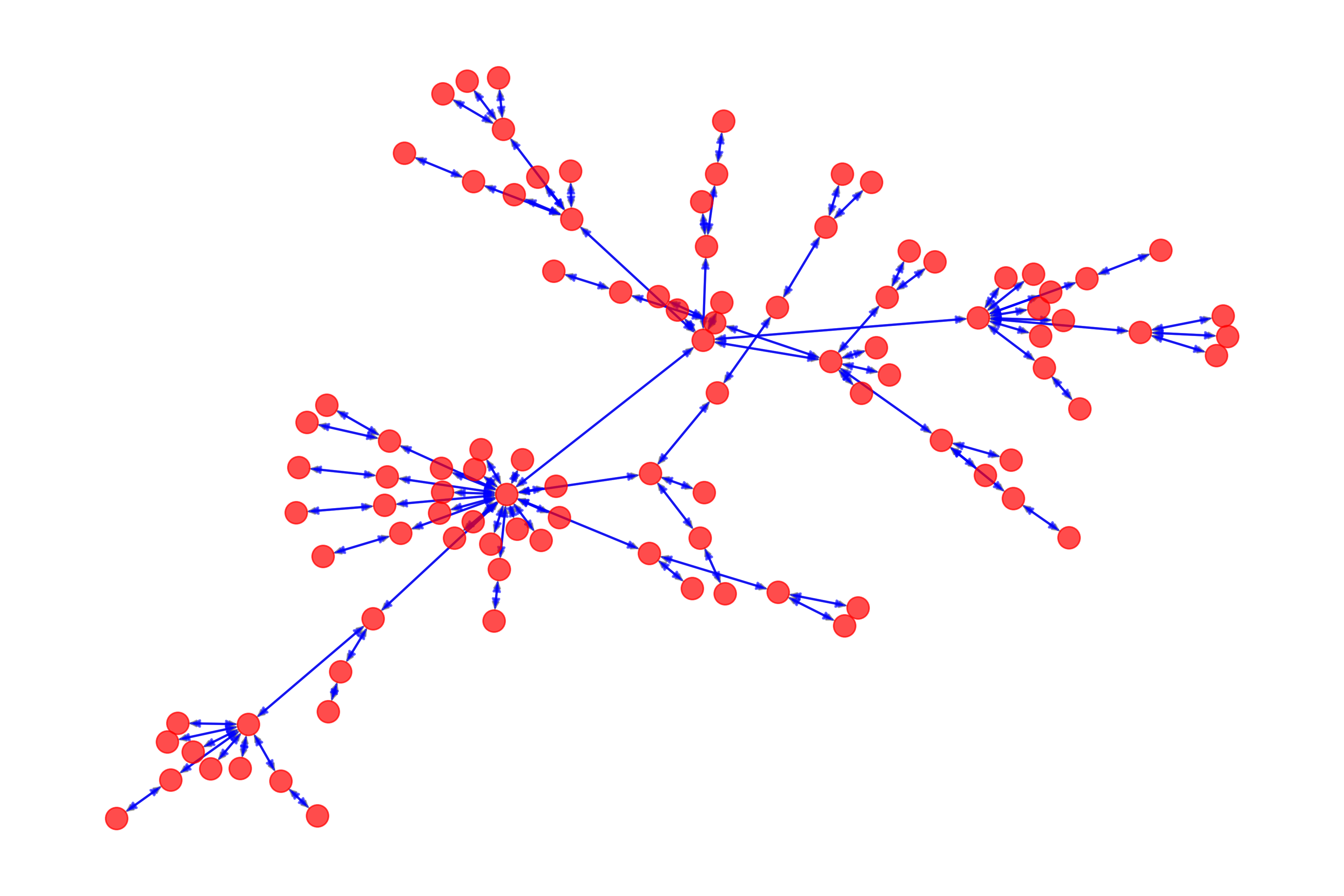}
    \hfill
    \includegraphics[width=0.23\textwidth]{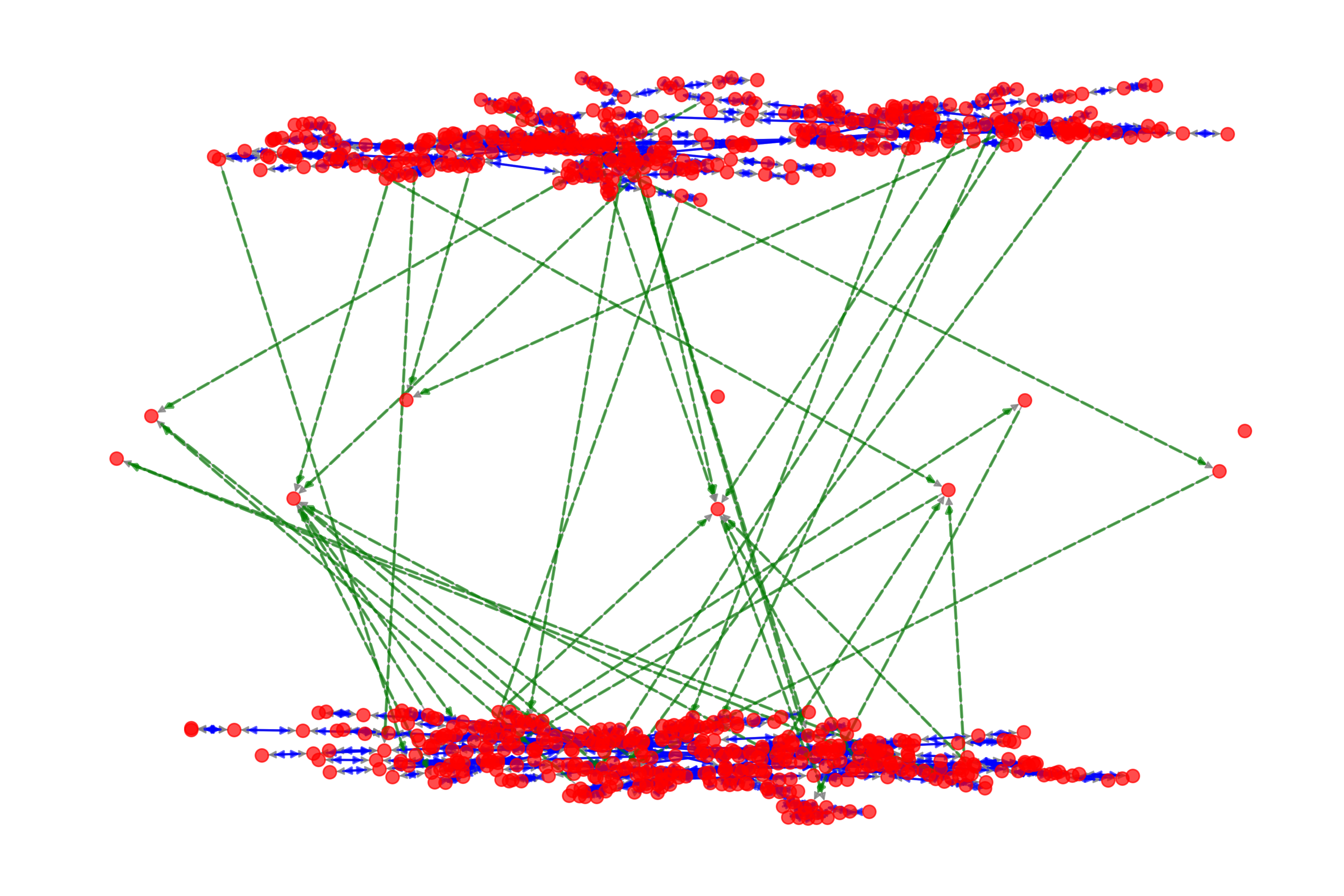}
    \hfill
    \includegraphics[width=0.23\textwidth]{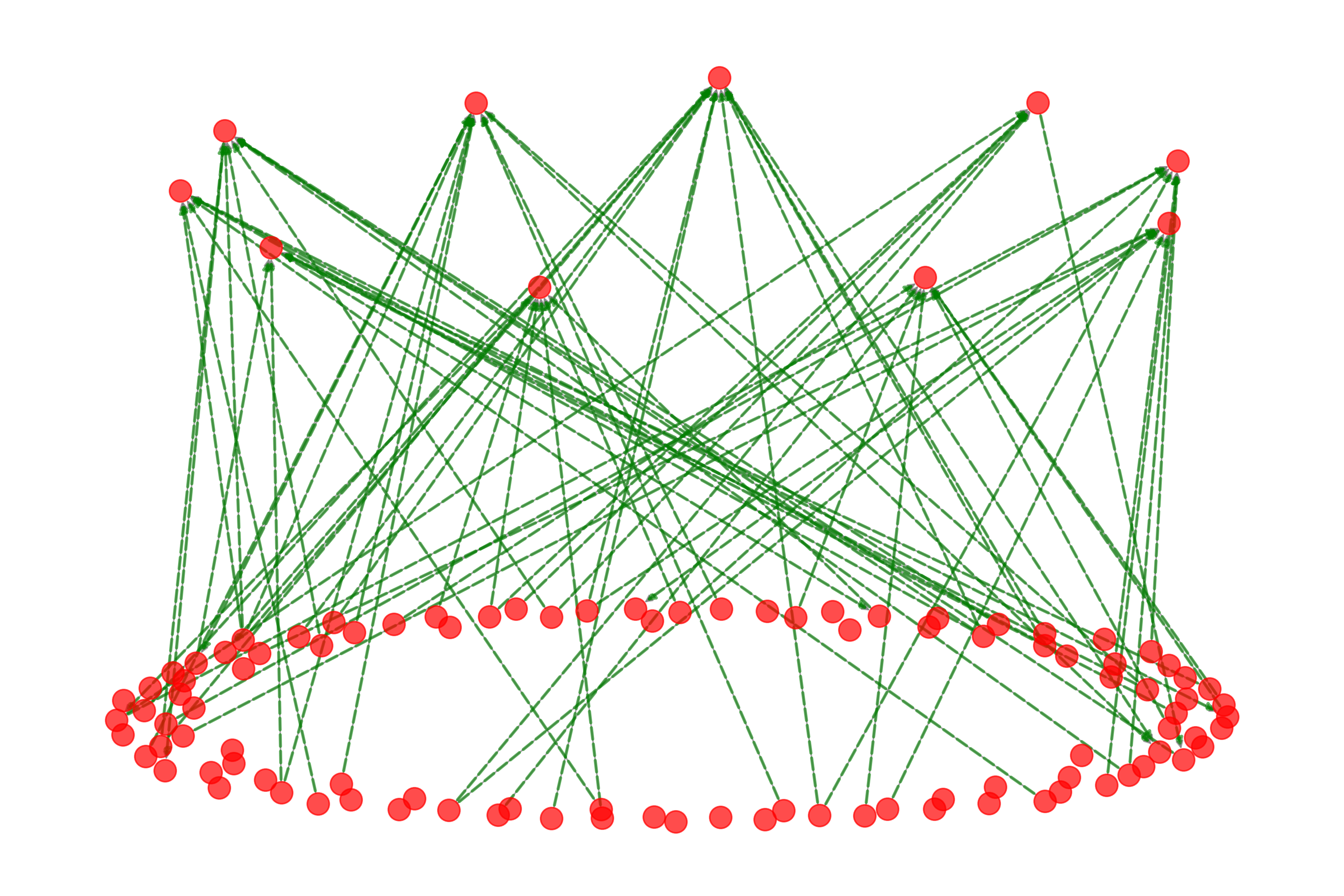}
    \caption{Network structures for TikTok (left-most), Twitter, Facebook and Reddit (right-most)}
    \label{fig:platform_architectures}
\end{figure*}

To create the network, we sampled from an actual Twitter follower network, consisting of approximately 41.65 million users and 14.7 billion edges~\cite{snap_dataset}. Specifically, to get a representative sample from the original network, we used random-walk sampling with restarts~\cite{leskovec2006sampling}. For comparison purposes, we sampled from the same network for Facebook's user--user and page--page networks (see Fig.~\ref{fig:platform_architectures}). 

As of the writing of this paper, Facebook has billions of daily active users, tens of millions of active groups, and hundreds of millions of pages. To maintain the rough proportions of users between users, groups, and pages, we ran our simulations for Facebook with 9,000 users, 1,000 pages and 100 groups. (Groups do not generate content, and therefore do not count towards the 10,000 agent limit.) Similarly, Reddit currently has hundreds of thousands of active subreddits and hundreds of millions of daily users. For consistency with Facebook we pick 100 subreddits for our model. We also limit the maximum number of groups, pages, and subreddits that a user can join at 10 of each type.

We modeled TikTok's network as a fully connected graph, where each user is connected to every other user, resulting in $(n \times (n - 1))$ edges for a network of $n$ users. In summary, our network had 10,000 users with 99,990,000 edges for TikTok, 10,000 users with 2,281,390 edges for Twitter, 9,000 users, 1,000 pages and 100 groups with 1,839,636 edges for Facebook, and 10,000 users with 100 groups with 2,000 edges for Reddit. Our setup for running the simulations is an Intel(R) Xeon(R) Gold 6338 CPU @ 2.00GHz with 251GB RAM. 


\subsection*{Statistical Analysis}
We fitted six hurdle models to characterize the effects of platform architecture and recommendation algorithm on engagement, one binary (which we refer to as `breadth') and one count model (`depth') for each of three outcomes: reach (whether a message was included in the feed of a user, regardless of exposure), exposure (whether a message was seen by an agent and how many agents it reached), resharing (whether any agent reshared a message and how many times it was reshared), and liking (whether any agent liked a message and how many likes it received). Binary outcomes were modeled using mixed-effects logistic regression (binomial family, logit link). Count outcomes were modeled using log-normal regression (Gaussian family, identity link on log-transformed counts) restricted to messages with at least one engagement event, consistent with the hurdle model framework. All models included platform architecture (Reddit, Facebook, Twitter, TikTok; reference: Reddit), recommendation algorithm (LIFO, Hot; reference: LIFO), their interaction, and content quality dimensions (illuminating, motivating) as fixed effects, with a random intercept for simulation seed to account for between-run variability. Model fit for count models was assessed using DHARMa nonparametric dispersion tests; all six models showed excellent fit (dispersion $\approx$ 1.000, all $p > .93$). All models were fitted in R using glmmTMB version 1.1.9.

To establish whether platform architecture and algorithm causally affect the quality of content agents encounter within our model, we fitted two encounter-level Gaussian mixed-effects models predicting message quality at the agent-message level, using all agent-message encounters in the exposures. The first model (total effect) included platform architecture, recommendation algorithm, their interaction, and agent quality preference as fixed effects, with random intercepts for message nested within seed to account for the non-independence of multiple agents encountering the same message. The second model additionally included content composition covariates (illuminating \& motivating) to decompose the direct and composition-mediated effects of platform and algorithm on encountered content quality. 


\subsubsection*{Artificial Intelligence}
The authors used Claude Sonnet 4.6 (Anthropic, San Francisco, CA) to assist with statistical analysis, code development, figure design, and manuscript preparation. All scientific content, interpretations, and conclusions are the responsibility of the authors.



\newcommand{\acknow}[1]{\paragraph{Acknowledgments.} #1}

\acknow{This material is based on work supported in part by the Institute for Trustworthy AI in Law and Society (TRAILS), which is supported by National Institute of Standards and Technology and the National Science Foundation under Award No. 2229885. Any opinions, findings, and conclusions or recommendations expressed in this material are those of the author(s) and do not necessarily reflect the views of the National Science Foundation or the National Institute of Standards and Technology. DAB and JS acknowledge support from the John S. and James L. Knight Foundation via the GW Institute for Data, Democracy, and Politics. GLC acknowledges support from the NSF under CAREER grant no.~2239194 and from the University of Maryland Social Data Science Center. The authors would like to thank Haneen Al-Rashid for feedback and insightful discussions and Dr. Howie Huang for access to computational resources.}



\bibliographystyle{abbrv}
\bibliography{reference}

@Article{jhaver2023decentralizing,
  author     = {Jhaver, Shagun and Frey, Seth and Zhang, Amy X.},
  title      = {Decentralizing {Platform} {Power}: {A} {Design} {Space} of {Multi}-{Level} {Governance} in {Online} {Social} {Platforms}},
  doi        = {10.1177/20563051231207857},
  issn       = {2056-3051},
  language   = {EN},
  number     = {4},
  pages      = {20563051231207857},
  url        = {https://doi.org/10.1177/20563051231207857},
  urldate    = {2025-12-04},
  volume     = {9},
  journal    = {Social Media + Society},
  month      = oct,
  publisher  = {SAGE Publications Ltd},
  shorttitle = {Decentralizing {Platform} {Power}},
  year       = {2023},
}

@Article{weng2012competition,
  author    = {Weng, Lilian and Flammini, Alessandro and Vespignani, Alessandro and Menczer, Fillipo},
  title     = {Competition among memes in a world with limited attention},
  number    = {1},
  pages     = {335},
  volume    = {2},
  journal   = {Scientific reports},
  publisher = {Nature Publishing Group UK London},
  year      = {2012},
}

@Article{broniatowski2019illuminate,
  author    = {Broniatowski, David A. and Reyna, Valerie F.},
  title     = {To illuminate and motivate: a fuzzy-trace model of the spread of information online},
  doi       = {10.1007/s10588-019-09297-2},
  issn      = {1572-9346},
  number    = {4},
  pages     = {431--464},
  volume    = {26},
  journal   = {Computational and Mathematical Organization Theory},
  month     = aug,
  publisher = {Springer Science and Business Media LLC},
  year      = {2019},
}

@Article{broniatowski2024role,
  author    = {Broniatowski, David A. and Hosseini, Pedram and Porter, Ethan V. and Wood, Thomas J.},
  title     = {The role of mental representation in sharing misinformation online.},
  doi       = {10.1037/xap0000517},
  issn      = {1076-898X},
  number    = {4},
  pages     = {511--538},
  volume    = {30},
  journal   = {Journal of Experimental Psychology: Applied},
  month     = dec,
  publisher = {American Psychological Association (APA)},
  year      = {2024},
}

@Article{van_der_linden_how_2024,
  author    = {van der Linden, Sander},
  title     = {How influencers and algorithms mobilize propaganda—and distort reality},
  number    = {8029},
  volume    = {633},
  journal   = {Nature},
  publisher = {Springer Science and Business Media LLC},
  year      = {2024},
}

@Article{brady2023algorithm,
  author    = {Brady, William J. and Jackson, Joshua Conrad and Lindström, Björn and Crockett, M. J.},
  title     = {Algorithm-mediated social learning in online social networks},
  doi       = {10.1016/j.tics.2023.06.008},
  issn      = {1364-6613, 1879-307X},
  language  = {English},
  number    = {10},
  pages     = {947--960},
  url       = {https://www.cell.com/trends/cognitive-sciences/abstract/S1364-6613(23)00166-3},
  urldate   = {2024-03-11},
  volume    = {27},
  journal   = {Trends in Cognitive Sciences},
  month     = oct,
  pmid      = {37543440},
  publisher = {Elsevier},
  year      = {2023},
}

@Misc{Broniatowski_Moses_2014,
  author    = {Broniatowski, David A and Moses, Joel},
  title     = {Flexibility, complexity, and controllability in large scale systems},
  publisher = {Massachusetts Institute of Technology. Engineering Systems Division},
  year      = {2014},
}

@TechReport{moehring2025better,
  author      = {Moehring, Alex and Cooper, Alissa and Narayanan, Arvind and Ovadya, Aviv and Redmiles, Elissa and Allen, Jeff and Stray, Jonathan and Kamin, Julia and Sigerson, Leif and Thorburn, Luke and Motyl, Matt and Eslami, Motahhare and Johnson, Nadine Farid and Lubin, Nathaniel and Iyer, Ravi and Arnao, Zander},
  institution = {Knight--Georgetown Institute},
  title       = {Better Feeds: Algorithms That Put People First},
  type        = {resreport},
  url         = {https://kgi.georgetown.edu/wp-content/uploads/2025/02/Better-Feeds_-Algorithms-That-Put-People-First.pdf},
  month       = mar,
  year        = {2025},
}

@Article{ciampaglia_how_2018,
  author    = {Ciampaglia, Giovanni Luca and Nematzadeh, Azadeh and Menczer, Filippo and Flammini, Alessandro},
  title     = {How algorithmic popularity bias hinders or promotes quality},
  number    = {1},
  pages     = {15951},
  volume    = {8},
  journal   = {Scientific reports},
  publisher = {Nature Publishing Group UK London},
  year      = {2018},
}

@Article{wang_lower_2024,
  author    = {Wang, Stephanie and Huang, Shengchun and Zhou, Alvin and Metaxa, Dana{\"e}},
  title     = {Lower Quantity, Higher Quality: Auditing News Content and User Perceptions on Twitter/X Algorithmic versus Chronological Timelines},
  number    = {CSCW2},
  pages     = {1--25},
  volume    = {8},
  journal   = {Proceedings of the ACM on Human-Computer Interaction},
  publisher = {ACM New York, NY, USA},
  year      = {2024},
}

@Article{huszar2022algorithmic,
  author    = {Huszár, Ferenc and Ktena, Sofia Ira and O’Brien, Conor and Belli, Luca and Schlaikjer, Andrew and Hardt, Moritz},
  title     = {Algorithmic amplification of politics on {Twitter}},
  doi       = {10.1073/pnas.2025334119},
  number    = {1},
  pages     = {e2025334119},
  urldate   = {2022-07-05},
  volume    = {119},
  journal   = {Proceedings of the National Academy of Sciences},
  month     = jan,
  publisher = {Proceedings of the National Academy of Sciences},
  year      = {2022},
}

@Article{coleman_combining_2022,
  author    = {Bak-Coleman, Joseph B and Kennedy, Ian and Wack, Morgan and Beers, Andrew and Schafer, Joseph S and Spiro, Emma S and Starbird, Kate and West, Jevin D},
  title     = {Combining interventions to reduce the spread of viral misinformation},
  number    = {10},
  pages     = {1372--1380},
  volume    = {6},
  journal   = {Nature Human Behaviour},
  publisher = {Nature Publishing Group UK London},
  year      = {2022},
}

@Article{Broniatowski2022COVIDMisinformation,
  author  = {Broniatowski, David A. and Kerchner, Daniel and Farooq, Fouzia and Huang, Xiaolei and Jamison, Amelia M. and Dredze, Mark and Quinn, Sandra Crouse and Ayers, John W.},
  title   = {Twitter and Facebook posts about COVID-19 are less likely to spread misinformation compared to other health topics},
  doi     = {10.1371/journal.pone.0261768},
  number  = {1},
  pages   = {e0261768},
  url     = {https://doi.org/10.1371/journal.pone.0261768},
  volume  = {17},
  journal = {PLOS ONE},
  year    = {2022},
}

@Article{broniatowski_measuring_2016,
  author    = {Broniatowski, David A and Moses, Joel},
  title     = {Measuring flexibility, descriptive complexity, and rework potential in generic system architectures},
  number    = {3},
  pages     = {207--221},
  volume    = {19},
  journal   = {Systems Engineering},
  publisher = {Wiley Online Library},
  year      = {2016},
}

@Article{smaldino2025information,
  author    = {Smaldino, Paul E and Russell, Adam and Zefferman, Matthew R and Donath, Judith and Foster, Jacob G and Guilbeault, Douglas and Hilbert, Martin and Hobson, Elizabeth A and Lerman, Kristina and Miton, Helena and others},
  title     = {Information architectures: a framework for understanding socio-technical systems},
  number    = {1},
  pages     = {13},
  volume    = {2},
  journal   = {npj Complexity},
  publisher = {Nature Publishing Group UK London},
  year      = {2025},
}

@Article{broniatowski2016effective,
  author    = {Broniatowski, David A and Hilyard, Karen M and Dredze, Mark},
  title     = {Effective vaccine communication during the disneyland measles outbreak},
  number    = {28},
  pages     = {3225--3228},
  volume    = {34},
  journal   = {Vaccine},
  publisher = {Elsevier},
  year      = {2016},
}

@InProceedings{leskovec2006sampling,
  author    = {Leskovec, Jure and Faloutsos, Christos},
  booktitle = {Proceedings of the 12th ACM SIGKDD International Conference on Knowledge Discovery and Data Mining},
  title     = {Sampling from large graphs},
  doi       = {10.1145/1150402.1150479},
  isbn      = {1595933395},
  location  = {Philadelphia, PA, USA},
  pages     = {631–636},
  publisher = {Association for Computing Machinery},
  series    = {KDD '06},
  url       = {https://doi.org/10.1145/1150402.1150479},
  address   = {New York, NY, USA},
  numpages  = {6},
  year      = {2006},
}

@Article{broniatowski2025explaining,
  author    = {Broniatowski, David A. and Zhong, Wei and Simons, Joseph R. and Jamison, Amelia M. and Dredze, Mark and Abroms, Lorien C.},
  title     = {Explaining Twitter’s inability to effectively moderate content during the COVID-19 pandemic},
  doi       = {10.1038/s41598-025-20033-6},
  issn      = {2045-2322},
  number    = {1},
  volume    = {15},
  journal   = {Scientific Reports},
  month     = {oct},
  publisher = {Springer Science and Business Media LLC},
  year      = {2025},
}

@Misc{cima2025investigating,
  author        = {Lorenzo Cima and Benedetta Tessa and Stefano Cresci and Amaury Trujillo and Marco Avvenuti},
  title         = {Investigating the heterogenous effects of a massive content moderation intervention via Difference-in-Differences},
  doi           = {https://doi.org/10.1016/j.osnem.2025.100320},
  eprint        = {2411.04037},
  url           = {https://arxiv.org/abs/2411.04037},
  archiveprefix = {arXiv},
  primaryclass  = {cs.CY},
  year          = {2025},
}

@Article{zhang_form-_2024,
  author    = {Zhang, Amy X and Bernstein, Michael S and Karger, David R and Ackerman, Mark S},
  title     = {Form-from: A design space of social media systems},
  number    = {CSCW1},
  pages     = {1--47},
  volume    = {8},
  journal   = {Proceedings of the ACM on Human-Computer Interaction},
  publisher = {ACM New York, NY, USA},
  year      = {2024},
}

@InProceedings{kleppmann2024bluesky,
  author        = {Kleppmann, Martin and Frazee, Paul and Gold, Jake and Graber, Jay and Holmgren, Daniel and Ivy, Devin and Johnson, Jeromy and Newbold, Bryan and Volpert, Jaz},
  booktitle     = {Proceedings of the ACM Conext-2024 Workshop on the Decentralization of the Internet},
  title         = {Bluesky and the AT Protocol: Usable Decentralized Social Media},
  doi           = {10.1145/3694809.3700740},
  eprint        = {2402.03239},
  isbn          = {9798400712524},
  location      = {Los Angeles, CA, USA},
  pages         = {1–7},
  publisher     = {Association for Computing Machinery},
  series        = {DIN '24},
  url           = {https://doi.org/10.1145/3694809.3700740},
  address       = {New York, NY, USA},
  archiveprefix = {arXiv},
  copyright     = {Creative Commons Attribution 4.0 International},
  numpages      = {7},
  year          = {2024},
}

@Misc{snap_dataset,
  author       = {Jure Leskovec},
  title        = {Twitter follower network},
  howpublished = {https://snap.stanford.edu/data/twitter-2010.html},
  publisher    = {Stanford University},
  year         = {2010},
}

@Article{li2024platform,
  author    = {Li, Mengyu and Suk, Jiyoun and Zhang, Yini and Pevehouse, Jon C. and Sun, Yibing and Kwon, Hyerin and Lian, Ruixue and Wang, Rui and Dong, Xinxia and Shah, Dhavan V.},
  title     = {Platform affordances, discursive opportunities, and social media activism: A cross-platform analysis of \#{MeToo} on Twitter, Facebook, and Reddit, 2017–2020},
  doi       = {10.1177/14614448241285562},
  issn      = {1461-7315},
  number    = {1},
  pages     = {119--147},
  volume    = {28},
  journal   = {New Media \& Society},
  month     = oct,
  publisher = {SAGE Publications},
  year      = {2024},
}

@Article{piccardi2025reranking,
  author    = {Piccardi, Tiziano and Saveski, Martin and Jia, Chenyan and Hancock, Jeffrey and Tsai, Jeanne L. and Bernstein, Michael S.},
  title     = {Reranking partisan animosity in algorithmic social media feeds alters affective polarization},
  doi       = {10.1126/science.adu5584},
  note      = {arXiv:2411.14652 [cs]},
  number    = {6776},
  pages     = {eadu5584},
  url       = {https://www.science.org/doi/10.1126/science.adu5584},
  urldate   = {2025-12-02},
  volume    = {390},
  journal   = {Science},
  month     = nov,
  publisher = {American Association for the Advancement of Science},
  year      = {2025},
}

@Article{milli2025engagement,
  author   = {Milli, Smitha and Carroll, Micah and Wang, Yike and Pandey, Sashrika and Zhao, Sebastian and Dragan, Anca D},
  title    = {Engagement, user satisfaction, and the amplification of divisive content on social media},
  doi      = {10.1093/pnasnexus/pgaf062},
  eprint   = {https://academic.oup.com/pnasnexus/article-pdf/4/3/pgaf062/62267673/pgaf062.pdf},
  issn     = {2752-6542},
  number   = {3},
  pages    = {pgaf062},
  url      = {https://doi.org/10.1093/pnasnexus/pgaf062},
  volume   = {4},
  journal  = {PNAS Nexus},
  month    = {03},
  year     = {2025},
}

@TechReport{ovadya2023bridging,
  author        = {Ovadya, Aviv and Thorburn, Luke},
  institution   = {CoRR},
  title         = {Bridging Systems: Open Problems for Countering Destructive Divisiveness across Ranking, Recommenders, and Governance},
  doi           = {10.48550/arXiv.2301.09976},
  eprint        = {2301.09976},
  note          = {arXiv:2301.09976 [cs]},
  url           = {http://arxiv.org/abs/2301.09976},
  urldate       = {2025-10-29},
  archiveprefix = {arXiv},
  copyright     = {arXiv.org perpetual, non-exclusive license},
  month         = {jan},
  primaryclass  = {cs.SI},
  publisher     = {arXiv},
  year          = {2023},
}

@InProceedings{baumann_optimal_2024,
  author    = {Baumann, Fabian and Halpern, Daniel and Procaccia, Ariel D and Rahwan, Iyad and Shapira, Itai and W{\"u}thrich, Manuel},
  booktitle = {Proceedings of the ACM Web Conference 2024},
  title     = {Optimal engagement-diversity tradeoffs in social media},
  pages     = {288--299},
  year      = {2024},
}

@InProceedings{cao-etal-2024-toxicity,
  author    = {Cao, Yang Trista and Domingo, Lovely-Frances and Gilbert, Sarah and Mazurek, Michelle L. and Shilton, Katie and Daum{\'e} Iii, Hal},
  booktitle = {Proceedings of the 2024 Conference on Empirical Methods in Natural Language Processing},
  title     = {Toxicity Detection is {NOT} all you Need: Measuring the Gaps to Supporting Volunteer Content Moderators through a User-Centric Method},
  doi       = {10.18653/v1/2024.emnlp-main.209},
  editor    = {Al-Onaizan, Yaser and Bansal, Mohit and Chen, Yun-Nung},
  pages     = {3567--3587},
  publisher = {Association for Computational Linguistics},
  url       = {https://aclanthology.org/2024.emnlp-main.209/},
  address   = {Miami, Florida, USA},
  month     = nov,
  year      = {2024},
}

@TechReport{eckles2022algorithmic,
  author      = {Dean Eckles},
  institution = {SocArXiV},
  title       = {Algorithmic transparency and assessing effects of algorithmic ranking},
  doi         = {10.31235/osf.io/c8za6},
  month       = {apr},
  publisher   = {Center for Open Science},
  year        = {2022},
}

@Article{sasahara_social_2021,
  author    = {Sasahara, Kazutoshi and Chen, Wen and Peng, Hao and Ciampaglia, Giovanni Luca and Flammini, Alessandro and Menczer, Filippo},
  title     = {Social influence and unfollowing accelerate the emergence of echo chambers},
  number    = {1},
  pages     = {381--402},
  volume    = {4},
  journal   = {Journal of Computational Social Science},
  publisher = {Springer},
  year      = {2021},
}

@Article{orben2025fixing,
  author  = {Orben, Amy and Matias, J. Nathan},
  title   = {Fixing the science of digital technology harms},
  doi     = {10.1126/science.adt6807},
  number  = {6743},
  pages   = {152--155},
  url     = {https://www.science.org/doi/abs/10.1126/science.adt6807},
  volume  = {388},
  journal = {Science},
  year    = {2025},
}

@Article{wei_characterizing_2019,
  author    = {Wei, Zhenglin and Broniatowski, David A},
  title     = {Characterizing System Architectures Using Network Data},
  pages     = {301--308},
  volume    = {153},
  journal   = {Procedia Computer Science},
  publisher = {Elsevier},
  year      = {2019},
}

@Article{Broniatowski_etal2023,
  author    = {Broniatowski, David A and Simons, Joseph R and Gu, Jiayan and Jamison, Amelia M and Abroms, Lorien C},
  title     = {The efficacy of Facebook’s vaccine misinformation policies and architecture during the COVID-19 pandemic},
  number    = {37},
  pages     = {eadh2132},
  volume    = {9},
  journal   = {Science Advances},
  publisher = {American Association for the Advancement of Science},
  year      = {2023},
}

@Article{mccabe2024deplatforming,
  author  = {McCabe, Stefan D. and Ferrari, Diogo and Green, Jon and Lazer, David M. J. and Esterling, Kevin M.},
  title   = {Post-January 6th Deplatforming Reduced the Reach of Misinformation on Twitter},
  number  = {8015},
  pages   = {132--140},
  volume  = {630},
  journal = {Nature},
  year    = {2024},
}

\newpage

\IfStandalone{%
    \title{Supplementary Information}
    \date{}
    \maketitle
}{%
    \appendix
    \begin{center}
    \Large\textbf{Supplementary Information}
    \end{center}
}






\section{User characteristics}
\label{sec:si-user-characteristics}

A user/page $a\in\mathcal{A}$ is an agent described by the following traits:
\begin{itemize}
    \item Quality preference $\phi_a\in [-1, 1]$: A scalar reflecting the
    agent's preference for quality content. This value is sampled from a
    uniform distribution.
    %
    \item The reward sensitivity $\rho_a\in[0, 1]$ and meaning seeking $\mu_a\in[0, 1]$, which reflect the motivation to either seek engagement (reward sensitivity) and
    interpretive depth (meaning seeking) when scrolling through their feed. Both of these two traits are
    independently drawn from triangular distributions. These two traits are used to determine the likelihood to share content, which is described next.
    %
    \item  The probability to re-share messages $p^{(\mathrm{s})}_a$, which is defined as the vector norm $p^{(\mathrm{s})}_a=\|\boldsymbol\sigma_a\|_2=\sqrt{(1 - \rho_a)^2 + (1 - \mu_a)^2}$.
    %
    \item Let $p^{(\mathrm{l})}_{a,m}$ be the probability that user $a\in\mathcal{A}$ likes a message $m\in\mathcal{M}$, which is defined as the weighted average $p^{(\mathrm{l})}_{a,m} = (1 - \phi_a)k_m + \phi_a\beta_m$, where $k_m$ and $\beta_m$ are the message `clickbaitness' and informativeness, respectively, which are defined below.
\end{itemize}


\section{Message Characteristics}
\label{sec:si-message-characteristics}

A messages $m\in\mathcal{M}$ is described by the following traits:
\begin{enumerate}
    \item Motivational valence (i.e., degree of appeal or clickbait content) $k_m\in [0, 1]$, indicating how attention-grabbing the message is;

    \item Illuminating power $\beta_m\in [0, 1]$, indicating how meaningful or insightful the message is;

    \item Informativeness or \emph{quality} $\alpha_m \in [-1, 1]$, where a value of $-1$ indicates a misleading message, $0$ neutral, and $1$ and informative and factual message.
\end{enumerate}

These quantities are sampled for each message. Both $k_m$ and  $\beta_m$ are sampled from the same triangular distribution to model skewed preferences, such as most messages being moderately engaging but not highly insightful. The informativeness $\alpha_m$ is instead sampled uniformly at random and is used as the measure of content \emph{quality} in the main manuscript. Finally, the pair $(k_m, \beta_m)$ is used to calculate the \emph{message magnitude} used in the simulation loop. This is defined as $\sqrt{k_m^2 + \beta_m^2}$, providing a combined measure of the message's overall strength.

\section{Model simulation}
\label{sec:si-model-simulation}

To simulate the model, we implement a message-passing simulator over a graph $G$ whose topology depends on the particular architecture being simulated. Nodes refer to the various types of agents described in the main text for each architecture, and edges are connections among them. We describe the sampling algorithm with pseudocode given in Algorithm~\ref{alg:simulator}. 

At each time step, a random agent $a\in\mathcal{A}$ which could be either a user
or page (if present in the architecture), is activated. If groups or subreddits are present, the agent pulls all messages from the groups/subreddits that is part of. Then, the agent may produce a
new message with probability $p_{\rm post}$, or attempt to reshare with
probability $p_{\rm reshare}$. For reshares, candidate messages are drawn from
the agent's input queue. They are first filtered by their credibility (i.e.,
informativeness above the agent's quality preference) and share tendency
(minimum message magnitude). If any messages meet these criteria, one of them is
chosen at random and forwarded.

The simulator records all message-level events (creation, resharing, etc.) to
enable analysis of message diffusion under varying algorithmic and architectural
conditions, as described in the main text.

\begin{algorithm}[H]
\caption{Message Passing Simulator}
\label{alg:simulator}
\footnotesize
\begin{algorithmic}[1]

\Statex \textbf{Inputs:}
\Statex $\mathcal{A} = \{a_1, \ldots a_{N}\}$: set of posting/resharing agents (users or pages)
\Statex $\mathcal{S} = \{s_1, \ldots s_{M}\}$: set of non-posting agents (groups or subreddits)
\Statex $G = (V, E)$: where $V \defeq \mathcal{A} \mathbin{\dot{\cup}} \mathcal{S}$ Graph of agents (nodes) and connections (edges)
\Statex Let $\mathcal{N}: V \to 2^V$ denote the neighbors of an agent
\Statex Let $\mathcal{L}: \mathcal{M} \to \mathbb{N}$ denote the number of likes of each message
\Statex $T>0$: Number of simulation steps
\Statex $p_{\rm post}$: Probability an agent produces a message
\Statex $p_{\rm share}$: Probability an agent attempts to reshare a message

\State $\mathcal{M} \gets \emptyset$
\For{$t = 1, \ldots, T$:}

    \State $a \defeq \left(\phi_a, \rho_a, \mu_a \right) \gets$ random choice from $\mathcal{A}$
    \State \texttt{/* Pull messages from any group or subreddits, if any */}
    \If{$\mathcal{N}(a) \cap \mathcal{S} \neq \emptyset$}
        \State Let $\mathcal{N}_g(a) \defeq \mathcal{N}(a) \cap \mathcal{S}$
        \For{$k = 1, \ldots, |\mathcal{N}_g(a)|$}
            \State $Q^{\rm in}_a \gets Q^{\rm in}_a \cup Q_{g_k}$ where $g_k \in \mathcal{N}_g(a)$
        \EndFor
    \EndIf
    \State $m$ $\gets$ \texttt{None}

    \If{$a$ is a producer $\wedge{}\, \mathrm{random()} < p_{\rm post}$}
        \State $m \gets \left(k_m, \beta_m, \alpha_m\right)$
        \State $\mathcal{M} \gets \mathcal{M} \cup \{m\}$
        \State Let $\mathcal{L}(m) \defeq 0$
    
    \ElsIf{$a$ is a resharer $\wedge{}\, \mathrm{random()} < p_{\rm share}$}
        \State \texttt{/* Filter input queue for candidate messages */} 
        \State $\mathcal{M}_{\rm cand} \gets \{m \in Q^{\rm in}_{a} \quad | \quad \alpha_{m} >  \phi_a \, \wedge \, \sqrt{k^2_m + \beta^2_m} \geq p_a^{(\rm s)}\}$ 
        \If{$\mathcal{M_{\rm cand}} \neq \emptyset$}
            \State $m \gets$ random choice from $\mathcal{M_{\rm cand}}$
            \State $Q^{\rm in}_a \gets Q^{\rm in}_a \setminus \{ m \}$
        \EndIf
    \EndIf
    \If{$m \neq$ \texttt{None}}
        \State \texttt{/* Push message to all neighbors of $a$ who are users, if any */}
        \State Let $\mathcal{N}_u(a) \defeq \mathcal{N}(a) \cap \mathcal{A} \wedge a_j$ is a user
        \If{$\mathcal{N}_u(a) \neq \emptyset$}
            \For{$a_j \in \mathcal{N}_u(a)$}
                \State $Q^{\rm in}_{a_j} \gets Q^{\rm in}_{a_j} \cup \{ m \}$
            \EndFor
        \EndIf
        \State \texttt{/* Push message to a random group/subreddits of $a$, if any */}
        \If{$\mathcal{N}(a) \cap \mathcal{S} \neq \emptyset$}
            \State $s_j \gets$ random choice from $\mathcal{N}(a) \cap \mathcal{S}$
            \State $Q^{\rm in}_{s_j} \gets Q^{\rm in}_{s_j} \cup \{ m \}$
        \EndIf
    \EndIf
    \State \texttt{/* Like the message with highest like probability in the feed */}
    \State Let  $N = |Q_a^{\rm in}|$ and $L_a \defeq (m_1, \ldots, m_N)$, with $p^{(\rm l)}_{a, m_1} \ge p^{(\rm l)}_{a, m_2} \ge \ldots \ge p^{(\rm l)}_{a, m_{N}}$
    \State $\mathcal{L}(m_1) \gets \mathcal{L}(m_1) + 1$
\EndFor
\end{algorithmic}
\end{algorithm}

\section{Supplementary Results}
\label{sec:si-supp-results}

\subsection{Dataset Overview}
We report the details of the dataset in Table~\ref{tbl:overview}.
In our simulations, seeds are independent initializations.
We use two algorithms (namely): chronological (LIFO) and engagement-ranked (Hot) feed orderings.
\begin{table}[htbp]
\centering
\begin{threeparttable}
\caption{Dataset Overview}
\label{tbl:overview}
\begin{tabular}{ll}
  \toprule
  \textbf{Characteristic} & \textbf{Value} \\
  \midrule
  Total observations (message-level) & 357,088 \\
  Seeds                              & 10 \\
  Algorithm conditions               & 2 (LIFO, Hot) \\
  Platforms                          & 4 (Reddit, Facebook, TikTok, Twitter) \\
  \bottomrule
\end{tabular}
\end{threeparttable}
\end{table}

\subsection{Zero-Inflation}
We check for zero inflation in our dataset.
Table~\ref{tbl:zeroinfl} highlights that reshares and likes are heavily zero-inflated in the dataset.
94.6\% and 97.2\% of messages received no reshares or likes respectively.
Reach (i.e., message ending up in user's queue without explicit viewing required) shows comparable zero-inflation (94.5\% zeros), as did the exposure, i.e., messages that were actually viewed by user's prior to resharing or liking.

\begin{table}[htbp]
\centering
\begin{threeparttable}
\caption{Zero-Inflation of Outcome and Exposure Variables}
\label{tbl:zeroinfl}
\begin{tabular}{lrr}
  \toprule
  \textbf{Variable} & \textbf{$N$ (total)} & \textbf{\% Zero} \\
  \midrule
  Reshares         & 357,088 & 94.6 \\
  Likes            & 357,088 & 97.2 \\
  Reach    & 357,088 & 94.5 \\
  Exposure & 357,088 & 94.5 \\
  \bottomrule
\end{tabular}
\end{threeparttable}
\end{table}

\subsection{Construct validity}
We validate key constructs within our data by checking for: a) reshares without prior exposure, b) distinctness of exposure from agent's queues, and c) likes without prior exposure.
Table~\ref{tbl:validity} highlights that no messages were reshared without reach (0 cases with reshares $>$ 0 and reach = 0), consistent with the constraint that sharing requires viewing.
Next, 144 messages had exposures $>$ 0 but were not reshared which highlights that the exposure is a genuinely distinct measure rather than a deterministic function of resharing.
Lastly, no messages had more likes than exposures (0 cases).

\begin{table}[htbp]
\centering
\begin{threeparttable}
\caption{Construct Validity Test}
\label{tbl:validity}
\begin{tabular}{lr}
  \toprule
  \textbf{Check} & \textbf{Count} \\
  \midrule
  Reshares $> 0$ \& Reach $= 0$ & 0 \\[4pt]
  Exposure $> 0$ \& Reshares $= 0$ & 144 \\[4pt]
  Likes $>$ Exposure & 0\\
  \bottomrule
\end{tabular}
\begin{tablenotes}
\small
\item \textit{Note.} \textit{Reach} = number of agents who had message in their queue. \textit{Exposures} = Exposure (agents
who opened the message for deliberate review). The 144 messages with exposure but no reshares confirm that these two measures
are distinct.
\end{tablenotes}
\end{threeparttable}
\end{table}

\subsection{Reach and Exposure}
Table~\ref{tbl:seen2} and~\ref{tbl:exposures} highlight that reach values vary substantially across platforms and exposure is considerably smaller in magnitude as exposure captures deliberate message viewing rather than appearing in agent queues.

\begin{table}[htbp]
\centering
\begin{threeparttable}
\caption{Distribution of Reach by Platform, Conditional on Exposure $> 0$.}
\label{tbl:seen2}
\begin{tabular}{lrrr}
  \toprule
  \textbf{Platform} & \textbf{Median} & \textbf{95th Percentile} & \textbf{Maximum} \\
  \midrule
  Reddit   &       2 &       4 &     7 \\
  Facebook &     266 &   2,420 & 5,875 \\
  Twitter  &     663 &   2,907 & 6,185 \\
  TikTok   &  9,999 &  9,999 & 9,999 \\
  \bottomrule
\end{tabular}
\begin{tablenotes}
\small
\item \textit{Note.} The statistics are computed on the subset of messages with \texttt{reach} $> 0$.
\end{tablenotes}
\end{threeparttable}
\end{table}

\begin{table}[htbp]
\centering
\begin{threeparttable}
\caption{Distribution of Exposure by Platform, Conditional on Exposure $> 0$}
\label{tbl:exposures}
\begin{tabular}{lrrr}
  \toprule
  \textbf{Platform} & \textbf{Median} & \textbf{95th Percentile} & \textbf{Maximum} \\
  \midrule
  Reddit   &  1 &   3 &   5 \\
  Facebook &  9 & 188 & 3,031 \\
  Twitter  & 14 & 219 & 3,401 \\
  TikTok   & 13 &  27 & 6,378 \\
  \bottomrule
\end{tabular}
\begin{tablenotes}
\small
\item \textit{Note.} Statistics computed on the subset of messages with
\texttt{exposures} $> 0$..
\end{tablenotes}
\end{threeparttable}
\end{table}
\subsection{Message Reach}
Table~\ref{tbl:struct_rates} shows the proportion of messages receiving any reach (whether a message entered any agent's queue, independent of active review).

\begin{table}[htbp]
\centering
\begin{threeparttable}
\caption{Proportion of Messages Receiving Any Reach and Unconditional Mean Reach Per Message, by Platform and Algorithm Condition}
\label{tbl:struct_rates}
\begin{tabular}{llS[table-format=2.1]S[table-format=4.1]}
  \toprule
  \textbf{Platform} & \textbf{Algorithm} & \textbf{\% Any Reach} & \textbf{Unconditional Mean} \\
  \midrule
  Reddit   & LIFO & 0.8  &    0.02 \\
  Reddit   & Hot  & 0.8  &    0.02 \\
  Facebook & LIFO & 8.8  &   60.0  \\
  Facebook & Hot  & 5.0  &   28.8  \\
  Twitter  & LIFO & 7.5  &   74.2  \\
  Twitter  & Hot  & 3.5  &   32.1  \\
  TikTok   & LIFO & 16.9 & 1686.0  \\
  TikTok   & Hot  & 0.2  &   22.9  \\
  \bottomrule
\end{tabular}
\begin{minipage}{\linewidth}
\centering
\footnotesize
Notes: Unconditional mean = proportion receiving any reach $\times$ mean reach among exposed messages. Captures total structural information flow per message across all messages including those receiving no exposure.
\end{minipage}
\end{threeparttable}
\end{table}



\subsection{Selection into Exposure}
\label{sec:sel_exposure}
We estimate the probability that a message receives any exposure using logistic regression.
The exposure is determined by messages that were explicitly viewed by agents while they pick one for resharing or liking from their queue.
We set the outcome variable as binary, indicating whether a message is viewed at least once.
Overall, this model captures whether content is surfaced to users at all.

\begin{table}[htbp]
\centering
\caption{Logistic regression estimates of the probability that a message receives any exposure (at least one view)}
\label{tbl:sel_exposure}
\begin{tabular}{lcccc}
\toprule
Term & Estimate & Std. Error & z-value & p-value \\
\midrule
Intercept                    & $-7.312$*** & 0.062 & $-117.38$ & $<$0.001 \\
Hot algorithm                & $-0.000$    & 0.074 &      0.00 & 1.000    \\
Facebook                     &  $2.472$*** & 0.055 &     44.97 & $<$0.001 \\
TikTok                       &  $3.258$*** & 0.054 &     60.48 & $<$0.001 \\
Twitter                      &  $2.303$*** & 0.055 &     41.62 & $<$0.001 \\
Illuminating                 &  $2.311$*** & 0.039 &     58.96 & $<$0.001 \\
Motivating                   &  $2.327$*** & 0.039 &     59.26 & $<$0.001 \\
Hot $\times$ Facebook        & $-0.619$*** & 0.079 &     $-7.84$ & $<$0.001 \\
Hot $\times$ TikTok          & $-4.559$*** & 0.124 &    $-36.76$ & $<$0.001 \\
Hot $\times$ Twitter         & $-0.831$*** & 0.080 &    $-10.34$ & $<$0.001 \\
\bottomrule
\end{tabular}
\begin{minipage}{\linewidth}
\centering
\footnotesize
Notes: Reference category is Reddit platform with LIFO algorithm.\\
Significance levels: * $p<0.05$, ** $p<0.01$, *** $p<0.001$.
\end{minipage}
\end{table}

Table~\ref{tbl:sel_exposure} shows that the probability of receiving any exposure is low at baseline (intercept at -7.31), reflecting that a small amount of messages receive exposure.
We observe that the platform effects are large and positive, indicating that Facebook, TikTok, and Twitter increase exposure likelihood relative to Reddit.
Additionally, content features (illuminating, motivating) also strongly increase exposure.
The hot algorithm has no main effect for Reddit, highlighting that the algorithm does not effect exposure on Reddit.
However, we observe large negative interactions across platforms, especially TikTok highlighting that a smaller subset of messages get more exposure on other platforms and that algorithm plays a significant role on platforms other than Reddit.

\subsection{Amplification of Exposure}
We examine the quantity of exposure conditional on a message receiving at least one view.
The dependent variable is the logarithm of exposure counts, and the model is estimated from the Gaussian family.
This approach captures the amplification stage, where content that passes the initial selection threshold may receive varying levels of attention.

\begin{table}[htbp]
\centering
\caption{Gaussian regression on the log-transformed number of exposures, estimated for messages that receive at least one exposure}
\label{tbl:amp_exposure}
\begin{tabular}{lcccc}
\toprule
Term & Estimate & Std. Error & z-value & p-value \\
\midrule
Intercept                    & $-0.183$*   & 0.089 &  $-2.07$ & 0.039    \\
Hot algorithm                &  $0.002$    & 0.105 &     0.02 & 0.987    \\
Facebook                     &  $2.214$*** & 0.078 &    28.52 & $<$0.001 \\
TikTok                       &  $2.069$*** & 0.076 &    27.24 & $<$0.001 \\
Twitter                      &  $2.521$*** & 0.078 &    32.27 & $<$0.001 \\
Illuminating                 &  $0.399$*** & 0.054 &     7.44 & $<$0.001 \\
Motivating                   &  $0.343$*** & 0.053 &     6.42 & $<$0.001 \\
Hot $\times$ Facebook        & $-0.443$*** & 0.112 &  $-3.97$ & $<$0.001 \\
Hot $\times$ TikTok          &  $6.168$*** & 0.176 &    34.96 & $<$0.001 \\
Hot $\times$ Twitter         & $-0.455$*** & 0.114 &  $-4.00$ & $<$0.001 \\
\bottomrule
\end{tabular}
\begin{minipage}{\linewidth}
\centering
\footnotesize
Notes: Reference category is Reddit platform with LIFO algorithm.\\
Significance levels: * $p<0.05$, ** $p<0.01$, *** $p<0.001$.
\end{minipage}
\end{table}

Table~\ref{tbl:amp_exposure} highlights that conditional on exposure, platform effects are strongly positive and content features also increase reach.
The key result is the large positive TikTok and hot algorithm interaction.
This implies that while the hot algorithm suppresses exposure probability (as seen in Section~\ref{sec:sel_exposure}), it produces extreme amplification for the small subset of content.

\subsection{Structural Selection into Visibility}
We estimate the probability that a message has reach (i.e., ended up in the agent's queue), which might not be explicitly viewed by the agent for resharing or liking (i.e., exposure).
Similar to Section~\ref{sec:sel_exposure}, it is a logistic regression with a binary outcome.
Through this process, we assess whether the patterns observed in realized exposure reflect underlying platform design.

\begin{table}[htbp]
\centering
\caption{Logistic regression estimates of whether a message ended up in agent's queues independent of exposure}
\label{tbl:struct_sel}
\begin{tabular}{lcccc}
\toprule
Term & Estimate & Std. Error & z-value & p-value \\
\midrule
Intercept                    & $-7.312$*** & 0.062 & $-117.38$ & $<$0.001 \\
Hot algorithm                & $-0.000$    & 0.074 &      0.00 & 1.000    \\
Facebook                     &  $2.472$*** & 0.055 &     44.97 & $<$0.001 \\
TikTok                       &  $3.258$*** & 0.054 &     60.48 & $<$0.001 \\
Twitter                      &  $2.303$*** & 0.055 &     41.62 & $<$0.001 \\
Illuminating                 &  $2.311$*** & 0.039 &     58.96 & $<$0.001 \\
Motivating                   &  $2.327$*** & 0.039 &     59.26 & $<$0.001 \\
Hot $\times$ Facebook        & $-0.619$*** & 0.079 &  $-7.84$  & $<$0.001 \\
Hot $\times$ TikTok          & $-4.559$*** & 0.124 & $-36.76$  & $<$0.001 \\
Hot $\times$ Twitter         & $-0.831$*** & 0.080 & $-10.34$  & $<$0.001 \\
\bottomrule
\end{tabular}
\begin{minipage}{\linewidth}
\centering
\footnotesize
Notes: Reference category is Reddit platform with LIFO algorithm.\\
Significance levels: * $p<0.05$, ** $p<0.01$, *** $p<0.001$.
\end{minipage}
\end{table}

Table~\ref{tbl:struct_sel} shows that the model closely replicates the exposure binary results, showing that structural visibility and actual exposure are governed by similar processes.

\subsection{Amplification of Reach}
We examine the extent of reach conditional on a message receiving at least one queue entry, using a log-normal regression (Gaussian family on log-transformed counts). This model captures how deeply messages penetrate the system structurally, independent of whether agents actively review them.

\begin{table}[htbp]
\centering
\caption{Gaussian regression on the log-transformed number of reach, estimated for messages that receive at least one queue entry}
\label{tbl:amp_struct}
\begin{tabular}{lcccc}
\toprule
Term & Estimate & Std. Error & z-value & p-value \\
\midrule
Intercept                    &  $0.466$*** & 0.098 &    4.77  & $<$0.001 \\
Hot algorithm                & $-0.000$    & 0.116 &    0.00  & 0.999    \\
Facebook                     &  $4.503$*** & 0.086 &   52.52  & $<$0.001 \\
TikTok                       &  $8.420$*** & 0.084 &  100.34  & $<$0.001 \\
Twitter                      &  $5.286$*** & 0.086 &   61.25  & $<$0.001 \\
Illuminating                 &  $0.324$*** & 0.059 &    5.47  & $<$0.001 \\
Motivating                   &  $0.222$*** & 0.059 &    3.76  & $<$0.001 \\
Hot $\times$ Facebook        & $-0.632$*** & 0.123 &  $-5.13$ & $<$0.001 \\
Hot $\times$ TikTok          &  $0.053$    & 0.195 &    0.27  & 0.786    \\
Hot $\times$ Twitter         & $-0.634$*** & 0.125 &  $-5.05$ & $<$0.001 \\
\bottomrule
\end{tabular}
\begin{minipage}{\linewidth}
\centering
\footnotesize
Notes: Reference category is Reddit platform with LIFO algorithm. Random intercept for simulation seed included but not shown. 
Significance levels: * $p<0.05$, ** $p<0.01$, *** $p<0.001$.
\end{minipage}
\end{table}

\subsection{Selection into Resharing}
We predict whether a message receives any reshares using logistic regression.

\begin{table}[htbp]
\centering
\caption{Logistic regression estimates of the probability that a message receives at least one reshare}
\label{tbl:log_prob}
\begin{tabular}{lcccc}
\toprule
Term & Estimate & Std. Error & z-value & p-value \\
\midrule
Intercept                    & $-7.391$*** & 0.064 & $-115.36$ & $<$0.001 \\
Hot algorithm                & $-0.000$    & 0.077 &      0.00 & 1.000    \\
Facebook                     &  $2.535$*** & 0.057 &     44.54 & $<$0.001 \\
TikTok                       &  $3.335$*** & 0.056 &     59.76 & $<$0.001 \\
Twitter                      &  $2.379$*** & 0.057 &     41.59 & $<$0.001 \\
Illuminating                 &  $2.317$*** & 0.039 &     58.88 & $<$0.001 \\
Motivating                   &  $2.326$*** & 0.039 &     59.01 & $<$0.001 \\
Hot $\times$ Facebook        & $-0.625$*** & 0.082 &   $-7.65$ & $<$0.001 \\
Hot $\times$ TikTok          & $-4.559$*** & 0.126 &  $-36.27$ & $<$0.001 \\
Hot $\times$ Twitter         & $-0.831$*** & 0.083 &  $-10.02$ & $<$0.001 \\
\bottomrule
\end{tabular}
\begin{minipage}{\linewidth}
\centering
\footnotesize
Notes: Reference category is Reddit platform with LIFO algorithm.\\
Significance levels: * $p<0.05$, ** $p<0.01$, *** $p<0.001$.
\end{minipage}
\end{table}

Table~\ref{tbl:log_prob} highlights that the reshare probability follows the same structure as exposure suggesting that engagement behavior is highly related to visibility mechanisms.

\subsection{Amplification of Resharing}
We examine the number of reshares conditional on receiving at least one reshare.
The dependent variable is log-transformed, and the model captures the intensity of diffusion among content that has already begun to spread.

\begin{table}[htbp]
\centering
\caption{Gaussian regression on the log-transformed number of reshares, estimated for messages with at least one reshare}
\label{tbl:amp_reshare}
\begin{tabular}{lcccc}
\toprule
Term & Estimate & Std. Error & z-value & p-value \\
\midrule
Intercept                    & $-0.528$*** & 0.040 & $-13.08$ & $<$0.001 \\
Hot algorithm                & $-0.000$    & 0.048 &   $-0.01$ & 0.994   \\
Facebook                     &  $0.300$*** & 0.036 &     8.42 & $<$0.001 \\
TikTok                       &  $0.094$**  & 0.035 &     2.69 & 0.007    \\
Twitter                      &  $0.355$*** & 0.036 &     9.92 & $<$0.001 \\
Illuminating                 &  $0.527$*** & 0.024 &    22.10 & $<$0.001 \\
Motivating                   &  $0.525$*** & 0.024 &    22.11 & $<$0.001 \\
Hot $\times$ Facebook        &  $0.049$    & 0.051 &     0.96 & 0.339    \\
Hot $\times$ TikTok          &  $3.574$*** & 0.079 &    45.10 & $<$0.001 \\
Hot $\times$ Twitter         &  $0.050$    & 0.052 &     0.97 & 0.334    \\
\bottomrule
\end{tabular}
\begin{minipage}{\linewidth}
\centering
\footnotesize
Notes: Reference category is Reddit platform with LIFO algorithm.\\
Significance levels: * $p<0.05$, ** $p<0.01$, *** $p<0.001$.
\end{minipage}
\end{table}

Table~\ref{tbl:amp_reshare} highlights that platform effects are smaller but still positive.
Additionally, TikTok stands out such that the large positive interaction indicates extreme amplification of viral content.

\subsection{Selection into Likes}
We predict whether a message receives any like using logistic regression.

\begin{table}[htbp]
\centering
\caption{Logistic regression estimates of the probability that a message receives at least one like}
\label{tbl:sel_like}
\begin{tabular}{lcccc}
\toprule
Term & Estimate & Std. Error & z-value & p-value \\
\midrule
Intercept                    & $-10.060$*** & 0.151 & $-66.62$ & $<$0.001 \\
Hot algorithm                &  $-0.000$    & 0.202 &     0.00 & 1.000    \\
Facebook                     &   $3.904$*** & 0.145 &    26.97 & $<$0.001 \\
TikTok                       &   $4.454$*** & 0.144 &    30.90 & $<$0.001 \\
Twitter                      &   $3.785$*** & 0.145 &    26.12 & $<$0.001 \\
Illuminating                 &   $1.801$*** & 0.052 &    34.48 & $<$0.001 \\
Motivating                   &   $3.937$*** & 0.056 &    70.13 & $<$0.001 \\
Hot $\times$ Facebook        &  $-0.665$**  & 0.206 &   $-3.23$ & 0.001   \\
Hot $\times$ TikTok          &  $-4.178$*** & 0.238 &  $-17.55$ & $<$0.001 \\
Hot $\times$ Twitter         &  $-0.832$*** & 0.206 &   $-4.03$ & $<$0.001 \\
\bottomrule
\end{tabular}
\begin{minipage}{\linewidth}
\centering
\footnotesize
Notes: Reference category is Reddit platform with LIFO algorithm.\\
Significance levels: * $p<0.05$, ** $p<0.01$, *** $p<0.001$.
\end{minipage}
\end{table}

Table~\ref{tbl:sel_like} highlights that the probability of receiving any likes is more selective than exposure.
Motivating content has a particularly large effect, suggesting emotional engagement could play a key role.
The hot algorithm again suppresses participation broadly, especially on TikTok.

\subsection{Amplification of Likes}
We examine the number of likes conditional on receiving at least one like. 
The outcome is log-transformed, and the model captures the intensity of user engagement.

\begin{table}[htbp]
\centering
\caption{Gaussian regression on the log-transformed number of likes, estimated for messages receiving at least one like}
\label{tbl:amp_likes}
\begin{tabular}{lcccc}
\toprule
Term & Estimate & Std. Error & z-value & p-value \\
\midrule
Intercept                    & $-1.199$*** & 0.177 &  $-6.784$ & $<$0.001 \\
Hot algorithm                &  $0.000$    & 0.237 &     0.000 & 1.000    \\
Facebook                     &  $1.337$*** & 0.169 &     7.900 & $<$0.001 \\
TikTok                       &  $0.943$*** & 0.169 &     5.598 & $<$0.001 \\
Twitter                      &  $1.502$*** & 0.169 &     8.863 & $<$0.001 \\
Illuminating                 &  $0.185$**  & 0.060 &     3.077 & 0.002    \\
Motivating                   &  $1.853$*** & 0.064 &    28.845 & $<$0.001 \\
Hot $\times$ Facebook        &  $0.024$    & 0.240 &     0.100 & 0.920    \\
Hot $\times$ TikTok          &  $3.991$*** & 0.279 &    14.327 & $<$0.001 \\
Hot $\times$ Twitter         &  $0.038$    & 0.241 &     0.157 & 0.875    \\
\bottomrule
\end{tabular}
\begin{minipage}{\linewidth}
\centering
\footnotesize
Notes: Reference category is Reddit platform with LIFO algorithm.\\
Significance levels: * $p<0.05$, ** $p<0.01$, *** $p<0.001$.
\end{minipage}
\end{table}

Table~\ref{tbl:amp_likes} highlights that there are modest platform effects.
TikTok shows a large amplification effect under hot algorithm, confirming that viral engagement is highly concentrated.

\subsection{Exposure Contrasts}
We present contrasts for exposure to quantify how the Hot algorithm compares to LIFO within each platform, as well as how platforms differ from one another under a common baseline.
We use two models, the first provides us with the probability of receiving any exposure (binary model) and second model presents the amount of exposure conditional on being seen (count model).

\begin{table}[htbp]
\centering
\caption{Pairwise comparisons based on estimated marginal means. Binary contrasts reflect differences in receiving any exposure; count contrasts reflect differences in log exposure conditional on exposure.}
\label{tbl:exp_contrasts}
\begin{tabular}{lcccc}
\toprule
Contrast & Estimate & Std. Error & z-value & p-value \\
\midrule
\multicolumn{5}{l}{\textit{Algorithm (LIFO $-$ Hot)}} \\
Reddit (Binary)   &  0.000    & 0.074 &    0.00 & 1.000    \\
Facebook (Binary) &  0.619*** & 0.079 &   7.84  & $<$0.001 \\
TikTok (Binary)   &  4.559*** & 0.124 &  36.76  & $<$0.001 \\
Twitter (Binary)  &  0.831*** & 0.080 &  10.34  & $<$0.001 \\
\addlinespace
Reddit (Count)    & $-0.002$  & 0.105 & $-0.02$ & 0.987    \\
Facebook (Count)  &  0.443*** & 0.112 &   3.97  & $<$0.001 \\
TikTok (Count)    & $-6.168$*** & 0.176 & $-34.96$ & $<$0.001 \\
Twitter (Count)   &  0.455*** & 0.114 &   4.00  & $<$0.001 \\
\midrule
\multicolumn{5}{l}{\textit{Platform (LIFO baseline)}} \\
Reddit $-$ Facebook & $-2.472$*** & 0.055 & $-44.97$ & $<$0.001 \\
Reddit $-$ TikTok   & $-3.258$*** & 0.054 & $-60.48$ & $<$0.001 \\
Reddit $-$ Twitter  & $-2.303$*** & 0.055 & $-41.62$ & $<$0.001 \\
Facebook $-$ TikTok & $-0.786$*** & 0.021 & $-36.56$ & $<$0.001 \\
Facebook $-$ Twitter &  0.170***  & 0.025 &   6.81  & $<$0.001 \\
TikTok $-$ Twitter  &  0.956***  & 0.025 &  38.64  & $<$0.001 \\
\bottomrule
\end{tabular}
\begin{minipage}{\linewidth}
\centering
\footnotesize
Significance levels: * $p<0.05$, ** $p<0.01$, *** $p<0.001$.
\end{minipage}
\end{table}

Table~\ref{tbl:exp_contrasts} highlights that under LIFO relative to Hot, exposure probability is higher on Facebook, Twitter and especially TikTok, confirming that Hot strongly suppresses selection.
However, conditional on exposure, the sign reverses on TikTok.
The negative estimate for LIFO - Hot shows that Hot produces greater reach once content is selected.

\subsection{Reshare Contrasts}
We report contrasts for resharing behavior separating the probability of receiving any reshares from the number of reshares conditional on diffusion.
We use this method to assess how algorithmic ranking influences the initiation of diffusion as well as its eventual scale, and how these dynamics vary across platforms.
Table~\ref{tbl:reshare_cont} highlights that reshare contrasts closely mirror exposure.
The Hot algorithm reduces the probability of resharing across platforms but dramatically increases reshare counts on TikTok conditional on diffusion.
We also observe that amplification effects for Facebook and Twitter are not significant in the count model which highlights that viral dynamics are concentrated on TikTok.

\begin{table}[htbp]
\centering
\caption{Pairwise comparisons for resharing behavior. Binary contrasts reflect differences in the probability of any reshare; count contrasts reflect differences in log reshares conditional on resharing.}
\label{tbl:reshare_cont}
\begin{tabular}{lcccc}
\toprule
Contrast & Estimate & Std. Error & z-value & p-value \\
\midrule
\multicolumn{5}{l}{\textit{Algorithm (LIFO $-$ Hot)}} \\
Reddit (Binary)   &  0.000    & 0.077 &    0.00  & 1.000    \\
Facebook (Binary) &  0.625*** & 0.082 &    7.65  & $<$0.001 \\
TikTok (Binary)   &  4.559*** & 0.126 &   36.27  & $<$0.001 \\
Twitter (Binary)  &  0.831*** & 0.083 &   10.02  & $<$0.001 \\
\addlinespace
Reddit (Count)    &  0.000    & 0.048 &    0.00  & 0.994    \\
Facebook (Count)  & $-0.049$  & 0.051 & $-0.96$  & 0.339    \\
TikTok (Count)    & $-3.574$*** & 0.079 & $-45.10$ & $<$0.001 \\
Twitter (Count)   & $-0.050$  & 0.052 & $-0.97$  & 0.334    \\
\midrule
\multicolumn{5}{l}{\textit{Platform (LIFO baseline)}} \\
Reddit $-$ Facebook  & $-2.535$*** & 0.057 & $-44.54$ & $<$0.001 \\
Reddit $-$ TikTok    & $-3.335$*** & 0.056 & $-59.76$ & $<$0.001 \\
Reddit $-$ Twitter   & $-2.379$*** & 0.057 & $-41.59$ & $<$0.001 \\
Facebook $-$ TikTok  & $-0.800$*** & 0.022 & $-36.30$ & $<$0.001 \\
Facebook $-$ Twitter &  0.156***   & 0.025 &   6.24   & $<$0.001 \\
TikTok $-$ Twitter   &  0.956***   & 0.026 &  36.90   & $<$0.001 \\
\bottomrule
\end{tabular}
\begin{minipage}{\linewidth}
\centering
\footnotesize
Significance levels: * $p<0.05$, ** $p<0.01$, *** $p<0.001$.
\end{minipage}
\end{table}

\subsection{Like Contrasts}
We present contrasts for liking behavior, distinguishing between the probability of receiving any likes and the number of likes conditional on engagement. 
These comparisons provide a direct measure of how algorithms and platforms shape user reactions from initial interaction to the intensity of engagement.
Table~\ref{tbl:like_contr} highlights that the Hot algorithm reduces the likelihood of receiving any likes but strongly increases like counts on TikTok once engagement occurs.
We also observe that platform contrasts show that TikTok generates the highest engagement overall, followed by Twitter and Facebook, with Reddit again as a low baseline.

\begin{table}[htbp]
\centering
\caption{Pairwise comparisons for liking behavior. Binary contrasts reflect differences in the probability of receiving any likes; count contrasts reflect differences in log likes conditional on receiving at least one.}
\label{tbl:like_contr}
\begin{tabular}{lcccc}
\toprule
Contrast & Estimate & Std. Error & z-value & p-value \\
\midrule
\multicolumn{5}{l}{\textit{Algorithm (LIFO $-$ Hot)}} \\
Reddit (Binary)   &  0.000    & 0.202 &    0.00  & 1.000    \\
Facebook (Binary) &  0.665**  & 0.206 &    3.23  & 0.001    \\
TikTok (Binary)   &  4.178*** & 0.238 &   17.55  & $<$0.001 \\
Twitter (Binary)  &  0.832*** & 0.206 &    4.03  & $<$0.001 \\
\addlinespace
Reddit (Count)    &  0.000    & 0.237 &    0.00  & 1.000    \\
Facebook (Count)  & $-0.024$  & 0.240 & $-0.10$  & 0.920    \\
TikTok (Count)    & $-3.991$*** & 0.279 & $-14.327$ & $<$0.001 \\
Twitter (Count)   & $-0.038$  & 0.241 & $-0.16$  & 0.875    \\
\midrule
\multicolumn{5}{l}{\textit{Platform (LIFO baseline)}} \\
Reddit $-$ Facebook  & $-3.904$*** & 0.145 & $-26.97$ & $<$0.001 \\
Reddit $-$ TikTok    & $-4.454$*** & 0.144 & $-30.90$ & $<$0.001 \\
Reddit $-$ Twitter   & $-3.785$*** & 0.145 & $-26.12$ & $<$0.001 \\
Facebook $-$ TikTok  & $-0.550$*** & 0.040 & $-13.75$ & $<$0.001 \\
Facebook $-$ Twitter &  0.119**    & 0.042 &   2.83   & 0.005    \\
TikTok $-$ Twitter   &  0.669***   & 0.042 &  15.93   & $<$0.001 \\
\bottomrule
\end{tabular}
\begin{minipage}{\linewidth}
\centering
\footnotesize
Significance levels: * $p<0.05$, ** $p<0.01$, *** $p<0.001$.
\end{minipage}
\end{table}

\subsection{Encounter-Level Models of Information Quality}

To establish whether platform architecture and recommendation algorithm causally affect the quality of content agents encounter, we fitted two encounter-level Gaussian mixed-effects models predicting message quality at the agent-message level, using all agent-message encounters in the exposures. Both models included platform architecture (Reddit, Facebook, Twitter, TikTok; reference: Reddit), recommendation algorithm (LIFO, Hot; reference: LIFO), their interaction, and agent quality preference as fixed effects, with random intercepts for message nested within seed ($1 \mid \text{seed/msg\_id}$) to account for the non-independence of multiple agents encountering the same message. The total effect model included no content composition controls; the controlled model additionally included \textit{illuminating} and \textit{motivating} as covariates, allowing decomposition of the direct effect of platform and algorithm on encountered content quality from any indirect effect operating through differential surfacing of content types. Model fit for the controlled model was assessed using DHARMa nonparametric dispersion tests (dispersion $= 1.028$, $p = .672$), confirming adequate fit (see Table \ref{tab:val_encounter}).

The key comparison is the Hot $\times$ TikTok interaction term: $b = -0.336$ in the total effect model and $b = -0.305$ in the controlled model. Controlling for content composition attenuated the TikTok/Hot coefficient by only 9\%, indicating that 91\% of the quality collapse under TikTok/Hot is a direct effect of the algorithm operating within TikTok's unconstrained architecture, not an artifact of differential surfacing of illuminating or motivating content.

\begin{table}[htbp]
\centering
\caption{Encounter-level Gaussian mixed-effects models predicting message quality ($val$) at the agent-message level. Each row = one agent--message encounter. Both models include random intercepts for message nested within seed. The total effect model includes no content composition controls; the controlled model additionally includes \textit{illuminating} and \textit{motivating}. Reference category: Reddit platform, LIFO algorithm. The 9\% attenuation of the Hot $\times$ TikTok coefficient from total to controlled model indicates that 91\% of the TikTok/Hot quality collapse is a direct architectural effect.}
\label{tab:val_encounter}
\small
\begin{tabular}{lcccc}
\toprule
& \multicolumn{2}{c}{\textbf{Total effect}} & \multicolumn{2}{c}{\textbf{Controlled}} \\
\cmidrule(lr){2-3}\cmidrule(lr){4-5}
\textbf{Term} & $b$ & 95\% CI & $b$ & 95\% CI \\
\midrule
Intercept                         &  $0.341$***  & $[0.327,\ 0.356]$  &  $0.154$***  & $[0.139,\ 0.170]$  \\
Hot algorithm                     &  $0.000$     & $[-0.013,\ 0.013]$ &  $0.000$     & $[-0.013,\ 0.013]$ \\
Facebook                          & $-0.024$***  & $[-0.037,\ -0.011]$ & $-0.024$***  & $[-0.037,\ -0.011]$ \\
Twitter                           & $-0.014$*    & $[-0.027,\ -0.001]$ & $-0.010$     & $[-0.023,\ 0.003]$ \\
TikTok                            &  $0.039$***  & $[0.026,\ 0.052]$  &  $0.036$***  & $[0.023,\ 0.049]$  \\
Agent Quality Preference        &  $0.001$***  & $[0.001,\ 0.001]$  &  $0.001$***  & $[0.001,\ 0.001]$  \\
\textit{illuminating}             & ---          & ---                 &  $0.206$***  & $[0.202,\ 0.210]$  \\
\textit{motivating}               & ---          & ---                 &  $0.115$***  & $[0.111,\ 0.119]$  \\
Hot $\times$ Facebook             & $-0.003$     & $[-0.016,\ 0.010]$ & $-0.005$     & $[-0.018,\ 0.008]$ \\
Hot $\times$ Twitter              & $-0.043$***  & $[-0.055,\ -0.030]$ & $-0.039$***  & $[-0.052,\ -0.026]$ \\
Hot $\times$ TikTok               & $-0.336$***  & $[-0.349,\ -0.323]$ & $-0.305$***  & $[-0.318,\ -0.292]$ \\
\addlinespace
\multicolumn{5}{l}{\textit{Random effects (SD)}} \\
msg\_id:seed                      & \multicolumn{2}{c}{$0.469$} & \multicolumn{2}{c}{$0.473$} \\
seed                              & \multicolumn{2}{c}{$0.000$} & \multicolumn{2}{c}{$0.000$} \\
Residual                          & \multicolumn{2}{c}{$0.106$} & \multicolumn{2}{c}{$0.106$} \\
\addlinespace
\multicolumn{5}{l}{\textit{Model fit}} \\
AIC                               & \multicolumn{2}{c}{$-2{,}437{,}531$} & \multicolumn{2}{c}{$-2{,}447{,}608$} \\
DHARMa dispersion                 & \multicolumn{2}{c}{---} & \multicolumn{2}{c}{$1.028$, $p = .672$} \\
\bottomrule
\end{tabular}
\begin{minipage}{\linewidth}
\smallskip
\footnotesize
* $p < .05$; *** $p < .001$. Confidence intervals are 95\% Wald CIs. Random effects reported as standard deviations. The near-zero seed SD in both models indicates negligible between-run variability after accounting for message-level random effects. Negative AIC values arise because the Gaussian log-likelihood can exceed zero when residual variance is small relative to the outcome scale.
\end{minipage}
\end{table}

\section{Descriptive Statistics by Platform and Algorithm}
Table~\ref{tab:breadth_depth_full} presents Breadth and depth of information spread by platform, algorithm, and outcome.

\begin{table}[htbp]
\centering
\caption{Breadth and depth of information spread by platform, algorithm, and outcome. Breadth = proportion of messages receiving at least one engagement event from any agent. Depth = mean number of unique agents engaging with each message, among messages with any engagement, reported as both arithmetic mean ($M$) and geometric mean ($GM$): $M$ corresponds to the unconditional mean calculation (breadth $\times$ depth); $GM$ corresponds directly to the log-normal model coefficients and is the primary inferential statistic. Under LIFO, Twitter produced greater depth than Facebook across all four outcomes on both mean measures (reach: $GM = 433$ vs $197$; exposure: $GM = 16.0$ vs $11.6$; reshares: $GM = 1.55$ vs $1.45$; likes: $GM = 4.88$ vs $4.09$), consistent with Moses' prediction that Twitter's network architecture is more flexible than Facebook's layered hierarchy. The TikTok seen2 ceiling effect ($GM = 9{,}999$) reflects that virtually all messages that enter circulation reach every agent's queue structurally; the meaningful distinction between TikTok conditions lies in active review depth, which increases 447-fold under Hot ($GM = 4{,}563$ vs $10.2$).}
\label{tab:breadth_depth_full}
\small
\begin{tabular}{llrrrrrrrrrrrr}
\toprule
& & \multicolumn{3}{c}{\textbf{Reach}} & \multicolumn{3}{c}{\textbf{Exposure}} & \multicolumn{3}{c}{\textbf{Reshares}} & \multicolumn{3}{c}{\textbf{Likes}} \\
\cmidrule(lr){3-5}\cmidrule(lr){6-8}\cmidrule(lr){9-11}\cmidrule(lr){12-14}
\textbf{Platform} & \textbf{Alg.} & \textit{B} & $M$ & $GM$ & \textit{B} & $M$ & $GM$ & \textit{B} & $M$ & $GM$ & \textit{B} & $M$ & $GM$ \\
\midrule
Reddit   & LIFO & 0.8\% & 2.3   & 2.19  & 0.8\% & 1.4   & 1.28  & 0.8\%  & 1.13 & 1.09 & 0.1\%  & 1.14 & 1.10 \\
Reddit   & Hot  & 0.8\% & 2.3   & 2.19  & 0.8\% & 1.4   & 1.28  & 0.8\%  & 1.13 & 1.09 & 0.1\%  & 1.14 & 1.10 \\
\addlinespace
Facebook & LIFO & 8.8\% & 683   & 197   & 8.8\% & 37.7  & 11.6  & 8.7\%  & 2.08 & 1.45 & 5.0\%  & 11.2 & 4.09 \\
Facebook & Hot  & 5.0\% & 574   & 104   & 5.0\% & 115   & 7.46  & 4.9\%  & 4.02 & 1.52 & 2.7\%  & 29.7 & 4.04 \\
\addlinespace
Twitter  & LIFO & 7.5\% & 987   & 433   & 7.5\% & 43.0  & 16.0  & 7.5\%  & 2.24 & 1.55 & 4.5\%  & 12.3 & 4.88 \\
Twitter  & Hot  & 3.5\% & 920   & 230   & 3.5\% & 180   & 10.1  & 3.5\%  & 5.28 & 1.63 & 2.0\%  & 41.0 & 4.80 \\
\addlinespace
TikTok   & LIFO & 16.9\% & 9,999 & 9,999 & 16.9\% & 13.1 & 10.2  & 16.9\% & 1.31 & 1.21 & 8.1\%  & 4.10 & 3.02 \\
TikTok   & Hot  & 0.2\% & 9,999 & 9,999 & 0.2\% & 5,807 & 4,563 & 0.2\%  & 92.5 & 39.0 & 0.1\%  & 959  & 129  \\
\bottomrule
\end{tabular}
\begin{minipage}{\linewidth}
\smallskip
\footnotesize
\textit{B} = breadth (proportion of all messages receiving any engagement from at least one agent). $M$ = arithmetic mean depth; $GM$ = geometric mean depth. Both depth measures computed among messages with at least one engagement event. Reach (seen2) = number of unique agents whose queue contained the message. Exposure (seen3) = number of unique agents who reviewed the message in their active review pool. Reshares and Likes = number of unique agents who reshared or liked the message. TikTok seen2 values of 9,999 reflect a ceiling effect: the agent population is 10,000 and virtually all messages that enter circulation reach every agent's queue structurally; the meaningful distinction between TikTok conditions lies in active review depth (seen3).
\end{minipage}
\end{table}

\end{document}